\pdfoutput=1

\documentclass[numberedappendix,twocolappendix,apj]{emulateapj}
\usepackage{apjfonts}
\usepackage[colorlinks=true,urlcolor=blue,linkcolor=blue,citecolor=blue]{hyperref}
\usepackage{amsmath}
\usepackage{natbib}
\usepackage{graphicx}

\usepackage{etoolbox}
\makeatletter
\patchcmd{\NAT@citex}
  {\@citea\NAT@hyper@{\NAT@nmfmt{\NAT@nm}\NAT@date}}
  {\@citea\NAT@nmfmt{\NAT@nm}\NAT@hyper@{\NAT@date}}
  {}% Do nothing if patch works
  {}% Do nothing if patch fails
\patchcmd{\NAT@citex}
  {\@citea\NAT@hyper@{%
     \NAT@nmfmt{\NAT@nm}%
     \hyper@natlinkbreak{\NAT@aysep\NAT@spacechar}{\@citeb\@extra@b@citeb}%
     \NAT@date}}
  {\@citea\NAT@nmfmt{\NAT@nm}%
   \NAT@aysep\NAT@spacechar%
   \NAT@hyper@{\NAT@date}}
  {}% Do nothing if patch works
  {}% Do nothing if patch fails
\patchcmd{\NAT@citex}
  {\@citea\NAT@hyper@{%
     \NAT@nmfmt{\NAT@nm}%
     \hyper@natlinkbreak{\NAT@spacechar\NAT@@open\if*#1*\else#1\NAT@spacechar\fi}%
       {\@citeb\@extra@b@citeb}%
     \NAT@date}}
  {\@citea\NAT@nmfmt{\NAT@nm}%
   \NAT@spacechar\NAT@@open\if*#1*\else#1\NAT@spacechar\fi%
   \NAT@hyper@{\NAT@date}}
  {}% Do nothing if patch works
  {}% Do nothing if patch fails
\makeatother

\graphicspath{{./}}

\def\MH2{M_{\rm H_2}}

\def\SFR{\dot{M}_\star}

\def\Fmdyn{F_{\rm -,d}}
\def\Fmfb{F_{\rm -,fb}}

\def\rhoSFR{{\dot{\rho}}_\star}
\def\rhoH2{\rho_{\rm H_2}}
\def\epsff{\epsilon_{\rm ff}}

\def\epsint{\epsilon}
\def\avir{\alpha_{\rm vir}}
\def\avirsf{\alpha_{\rm vir,sf}}
\def\cs{c_{\rm s}}
\def\st{\sigma_{\rm t}}
\def\stot{\sigma_{\rm tot}}
\def\rhosf{\rho_{\rm sf}}
\def\nsf{n_{\rm sf}}

\def\Sg{\Sigma_{\rm g}}

\def\SH2{\Sigma_{\rm H_2}}
\def\Ssf{\Sigma_{\rm sf}}

\def\SSFR{\dot\Sigma_\star}

\def\fsf{f_{\rm sf,H_2}}
\def\fH2{f_{\rm H_2}}

\def\Nc{N_{\rm c}}
\def\tsf{t_{\rm sf}}
\def\tnsf{t_{\rm nsf}}

\def\tglob{\tau}
\def\tH2{\tau_{\rm H_2}}

\def\tdep{t_\star}
\def\taust{\tau_\star}

\def\tmd{\tau_{\rm -,d}}

\def\tff{t_{\rm ff}}

\def\tmH2{t_{\rm H_2}}

\def\tmolnsf{t_{\rm H_2,nsf}}

\def\sigtot{\sigma_{\rm tot}}

\def\pc{{\rm\;pc}}
\def\kpc{{\rm\;kpc}}
\def\Myr{{\rm\;Myr}}
\def\Gyr{{\rm\;Gyr}}
\def\Msun{{\rm\;M_\odot}}
\def\Msunyr{{\rm\;M_\odot\;yr^{-1}}}
\def\Msunpc2{{\rm\;M_\odot\;pc^{-2}}}
\def\kms{{\rm\;km\;s^{-1}}}
\def\cc{{\rm\;cm^{-3}}}
\def\mpcc{\;m_{\rm p}{\rm\;cm^{-3}}}

\def\ns{$n$--$\sigtot$ }
\def\const{{\rm const}}
\def\Lstar{$L_\star$}

\def\q{\mathbf{q}}

\begin{document}

%%%%%%%%%%%%%%%%%%%%%%%%%%%%%%%%%%%%%%%%%%%%%%%%%%%%%%%%%%%%%%%%%%%%%
% HEADER:
%%%%%%%%%%%%%%%%%%%%%%%%%%%%%%%%%%%%%%%%%%%%%%%%%%%%%%%%%%%%%%%%%%%%%

\shorttitle{What sets the slope of the $\SSFR$--$\SH2$ relation?}
\shortauthors{Semenov, Kravtsov, Gnedin}
\slugcomment{Accepted for publication in the Astrophysical Journal}

\title{What Sets the Slope of the Molecular Kennicutt--Schmidt Relation?}

\author{Vadim A. Semenov\altaffilmark{1,2,$\star$}, Andrey V. Kravtsov\altaffilmark{1,2,3} and Nickolay Y. Gnedin\altaffilmark{1,2,4}}


\altaffiltext{1}{Department of Astronomy \& Astrophysics, The University of Chicago, Chicago, IL 60637 USA}
\altaffiltext{2}{Kavli Institute for Cosmological Physics, The University of Chicago, Chicago, IL 60637 USA}
\altaffiltext{3}{Enrico Fermi Institute, The University of Chicago, Chicago, IL 60637 USA}
\altaffiltext{4}{Fermilab Center for Particle Astrophysics, Fermi National Accelerator Laboratory, Batavia, IL 60510-0500 USA}
\altaffiltext{$\star$}{semenov@uchicago.edu}

%%%%%%%%%%%%%%%%%%%%%%%%%%%%%%%%%%%%%%%%%%%%%%%%%%%%%%%%%%%%%%%%%%%%%
% ABSTRACT:
%%%%%%%%%%%%%%%%%%%%%%%%%%%%%%%%%%%%%%%%%%%%%%%%%%%%%%%%%%%%%%%%%%%%%

\begin{abstract}
The surface densities of molecular gas, $\SH2$, and the star formation rate (SFR), $\SSFR$, correlate almost linearly on kiloparsec scales in observed star-forming (non-starburst) galaxies. We explore the origin of the linear slope of this correlation using a suite of isolated \Lstar~galaxy simulations. We show that in simulations with efficient feedback, the slope of the $\SSFR$--$\SH2$ relation on kiloparsec scales is insensitive to the slope of the \mbox{$\rhoSFR$--$\rho$} relation assumed at the resolution scale. We also find that the slope on kiloparsec scales depends on the criteria used to identify star-forming gas, with a linear slope arising in simulations that identify star-forming gas using a virial parameter threshold. This behavior can be understood using a simple theoretical model based on conservation of interstellar gas mass as the gas cycles between atomic, molecular, and star-forming states under the influence of feedback and dynamical processes. In particular, we show that the linear slope emerges when feedback efficiently regulates and stirs the evolution of dense, molecular gas. We show that the model also provides insights into the likely origin of the relation between the SFR and molecular gas in real galaxies on different scales.
\end{abstract} 

%--------------------------------------------------------------------
\section{Introduction}
\label{sec:intro}
%--------------------------------------------------------------------
\setcounter{footnote}{0}

One of the keys to understanding galaxy formation is understanding how galaxies form stars. Despite the complexity of the processes involved in star formation on $\lesssim 10$ parsec scales, observed star formation rates (SFRs) averaged on kiloparsec and larger scales exhibit a rather universal scaling with the amount of gas. This scaling is known as the Kennicutt--Schmidt relation \citep[KSR;][see also \citealt{Sanduleak.1969}, \citealt{Madore.etal.1974}]{Schmidt.1959,Schmidt.1963,Kennicutt.1989,Kennicutt.1998}, and it becomes especially tight and close to linear when only molecular gas is included in the estimate of the gas mass \citep[][]{Wong.Blitz.2002,Bigiel.etal.2008,Bigiel.etal.2011,Leroy.etal.2008,Leroy.etal.2013,Bolatto.etal.2017,Utomo.etal.2017,Colombo.etal.2018}. In other words, the depletion time of molecular gas in kiloparsec-scale patches of the interstellar medium (ISM),
\begin{equation}
\tH2 \equiv \frac{\SH2}{\SSFR}\approx 2\pm 1\ \mathrm{Gyr},
\end{equation}
is independent of the molecular gas surface density $\SH2$.

Although both total gas and molecular KSR were studied extensively in numerical simulations of galaxies \cite[e.g.,][]{Kravtsov.2003,Li.etal.2005,Saitoh.etal.2008,Schaye.DallaVecchia.2008,Feldmann.etal.2011,Gnedin.Kravtsov.2011,Rahimi.Kawata.2012,Agertz.etal.2013,Gnedin.etal.2014,Agertz.Kravtsov.2015,Khoperskov.Vasiliev.2017,Capelo.etal.2018,Lupi.etal.2018,Orr.etal.2018}, the physical origin of their normalization and scatter were understood only recently.
The normalization is directly related to the global depletion time in galaxies, which is long due to multiple cycles between star-forming and non-star-forming states that the ISM gas undergoes under the influence of  stellar feedback and dynamical processes \citep{Semenov.etal.2017}. Important sources of the scatter include intrinsic variations and decoupled evolution of molecular mass and SFR tracers in ISM patches \citep{Feldmann.etal.2011}, biased sampling of different parts of the gas cycle in such patches \citep{Kruijssen.Longmore.2014}, and non-equilibrium states of such patches \citep{Semenov.etal.2017}. 

The aspect of the molecular KSR that is still debated is its slope. The existence of a correlation between SFR and molecular gas surface densities is expected because both SFR and molecular gas trace dense ISM gas. What is surprising, however, is that the correlation observed in normal star-forming (non-starburst) galaxies is close to linear. Given that  dynamical timescales as $t_{\rm dyn} \propto 1/\sqrt{G\rho}$, where $\rho$ is the average density of a region, na\"ively we could expect a superlinear KSR: $\rhoSFR \propto \rho/t_{\rm dyn} \propto \rho^{1.5}$. Therefore, the linear slope indicates that the origin of molecular KSR is more nuanced and is not shaped by gas self-gravity alone. 

The mechanism responsible for the linear slope must be rather universal and operate in a range of environments and for different states of dense gas. Indeed, the near-linear slope in the molecular KSR persists in diverse galactic environments, from the average ISM of various Hubble types of disk galaxies \citep[e.g.,][]{Utomo.etal.2017,Colombo.etal.2018} to low-density disk outskirts \citep{Schruba.etal.2011} and even in low-metallicity dwarf galaxies \citep{Bolatto.etal.2011,Jameson.etal.2016}. A close-to-linear relation is also observed for dense gas visible in HCN and HCO$^+$ over 8 orders of magnitude in mass and on a wide range of spatial scales, from $\sim 10\pc$ scales to scales of entire galaxies \citep{Gao.Solomon.2004a,Gao.Solomon.2004b,Wu.etal.2005}. However, the relation for molecular gas is not always linear: it steepens on small, $\lesssim 10$ parsec, scales \citep[e.g.,][]{Evans.etal.2009,Evans.etal.2014,Heiderman.etal.2010,Gutermuth.etal.2011} and in extreme dense environments such as starburst galaxies \citep{Genzel.etal.2010,Genzel.etal.2015} and galactic centers \citep[e.g.,][]{Leroy.etal.2013}. Moreover, the depletion time exhibits mild trends with redshift and deviation of galaxies from the mean star formation sequence \citep[e.g.,][]{Tacconi.etal.2018}. To explain the origin of the molecular KSR slope, both its universality at moderate and low surface densities and its steepening in more extreme dense environments must be understood. Elucidating the origin of the linear slope will be the focus of this paper.

One of the most popular explanations for the close-to-linear slope of the molecular KSR is the so-called ``counting argument,'' which was first introduced by \citet{Wu.etal.2005} to interpret the linear relation observed for the dense gas traced by HCN and later extended to more diffuse molecular gas states. This argument posits that all molecular gas resides in star-forming units that have approximately the same properties and depletion times regardless of the galactic environment. In this case, both $\SH2$ and $\SSFR$ of an ISM patch result from counting these units in this patch, and therefore, $\SH2$ and $\SSFR$ become linearly related. However, the observed properties of molecular regions do vary with $\gtrsim 1\kpc$ galactic environment \citep[e.g.,][]{MivilleDeschenes.etal.2016}, and therefore, the explanation of the linear slope must be more nuanced.

Substantial progress in understanding the linear slope can be made by considering the timescales of gas evolution through different states rather than an instantaneous spatial realization of these states reflected in $\SSFR$ and $\SH2$. In the context of explaining KSR, such a ``dynamical'' point of view was first adopted by \citet{Madore.2010} and later extended by \citet{Elmegreen.2015,Elmegreen.2018}. Although this method is quite insightful, the linear slope in their models follows from rather strong assumptions that both the timescale of molecular gas evolution and the star formation efficiency on kiloparsec scales are constant and independent of $\SH2$ (see Section~\ref{sec:disc:prev} for details). As we show below, in our simulations, they are not constant and depend on $\SH2$. Yet, the molecular KSR in our fiducial simulations has a near-linear slope. Thus, these models cannot explain the physics of the molecular KSR in our simulations. 

Nevertheless, the way of thinking about KSR as a result of an interplay among different processes with their own characteristic timescales is quite insightful. As noted above, such an approach allows us to understand both the normalization and the scatter of KSR \citep{Kruijssen.Longmore.2014,Semenov.etal.2017,Semenov.etal.2018}. In this paper, we will use this approach to gain insights about the physical factors that determine the slope of the molecular KSR. 

Specifically, we will explore the origin of the molecular KSR in a suite of isolated \Lstar~galaxy simulations. As we will show, in our fiducial simulations, a near-linear relation emerges on kiloparsec scales and is insensitive to the slope adopted locally in the subgrid prescription for star formation. This insensitivity is a counterexample to the argument that in simulations, the KSR on kiloparsec scales simply reflects the relation assumed at the resolution scale \citep{Schaye.DallaVecchia.2008,Gnedin.etal.2014}. We show instead that the linear slope of the molecular KSR on kiloparsec scales and its insensitivity to small-scale star formation parameters is a result of self-regulation by stellar feedback.

This paper is organized as follows. In Section~\ref{sec:sims}, we briefly summarize our simulation suite. In Section~\ref{sec:results}, we demonstrate how the molecular KSR obtained in our simulations changes depending on the parameters of star formation and feedback. In Section~\ref{sec:pdf}, we show how the obtained results relate to the gas PDF in ISM patches and its scaling with gas surface density. In Section~\ref{sec:ksr}, we use our analytical model based on ISM gas cycling to explain our results and elucidate the origin of the linear slope of molecular KSR. In Section~\ref{sec:disc}, we compare our model to previous works and discuss its implications for the KSR derived in observations and simulations of galaxies. In Section~\ref{sec:summary}, we summarize our results and conclusions.

%--------------------------------------------------------------------
\section{Simulations}
\label{sec:sims}
%--------------------------------------------------------------------

We explore the origin of the molecular KSR slope using a suite of isolated $\sim$\Lstar~galaxy simulations with varying parameters of local star formation and feedback. Our simulation suite is detailed in \citet{Semenov.etal.2017,Semenov.etal.2018}, and here we briefly summarize the features that are most relevant for this study. 

The simulations are run using the adaptive mesh refinement gas dynamics and $N$-body code ART \citep{Kravtsov.1999,Kravtsov.etal.2002,Rudd.etal.2008,Gnedin.Kravtsov.2011}. We use the initial conditions from the AGORA code comparison project \citep{Kim.etal.2014,agora2}, which consist of gaseous and stellar disks and stellar bulge embedded in a dark matter halo. The galactic disk is exponential with the scale radius of $\approx 3.4\kpc$, scale height of $\approx 340\pc$, and total mass of $\approx 4.3\times 10^{10} \Msun$, $20\%$ of which is in the gaseous disk. The stellar bulge has a total mass of $\approx 4.3\times10^9 \Msun$ and a \citet{Hernquist.1990} density profile with the scale radius of $\approx 340\pc$. The dark matter halo has a Navarro--Frenk--White profile \citep{NFW.1996,NFW.1997} with the characteristic circular velocity of $v_{\rm c,200} = 150\kms$ and the concentration of $c=10$. Mesh cells are adaptively resolved when their gas mass exceeds $\sim 8\,300\Msun$ until the maximal resolution of $\Delta = 40\pc$ is reached.

To identify star-forming regions, we use results of theoretical models of star formation in a turbulent medium, which predict that star formation efficiency increases exponentially with decreasing local virial parameter $\avir$  \citep[for a review, see][and references therein]{Padoan.etal.2014}. For a computational cell with size $\Delta$, $\avir$ is defined as for a uniform sphere with mass $M$ and radius $R = \Delta/2$ \citep{Bertoldi.McKee.1992}:
\begin{equation}
\label{eq:avir}
\avir \equiv \frac{5 \stot^2 R}{3GM} \approx 9.35 \frac{ (\stot/10\kms)^2 }{ (n/100\cc) (\Delta/40 \pc)^2},
\end{equation}
where the total velocity dispersion $\stot = \sqrt{\cs^2 + \st^2}$ includes thermal velocity, $\cs$, and turbulent velocity, $\st$. In our simulations, we use the ``shear-improved'' subgrid turbulence model of \citet[][for details about the implementation in the ART code, see \citealt{Semenov.etal.2016}]{Schmidt.etal.2014}, which enables us to estimate $\st$ and $\stot$ in each computational cell. Although we can use the local estimate of $\avir$ to assign star formation efficiency to each cell, for simplicity of interpretation in this study we approximate the exponential dependence of the efficiency on $\avir$ using a sharp threshold $\avirsf$ and a constant efficiency for the gas that satisfies this threshold. Specifically, in our simulations, we define all cells with $\avir < \avirsf=10$ as star-forming, motivated by the fact that $\avir\lesssim10$ in observed molecular clouds \citep[e.g.,][]{Leroy.etal.2016,MivilleDeschenes.etal.2016}. For comparison, in some of the simulations, we will adopt a more commonly used gas density threshold, $\nsf$, instead of $\avir$, so that star-forming cells are identified as those with $n\geq\nsf$. 

In each cell satisfying the star formation threshold, the rate of star formation is assumed to be
\begin{equation}
\label{eq:rhoSFR-slope}
\rhoSFR = \epsff\, \frac{\rho_0}{t_{\rm ff,0}} \left( \frac{\rho}{\rho_0} \right)^{\beta},
\end{equation}
where $t_{\rm ff,0}=\sqrt{3 \pi / 32 G \rho_0}$ is the freefall time at $\rho_0 = 100\mpcc$, $\epsff$ is the star formation efficiency per freefall time assumed to be constant in the simulations discussed in this paper, and $\beta$ is the local slope of the SFR dependence on density. For $\beta=1.5$, this expression is equivalent to the most common choice since the first generation of simulations \citep{Cen.Ostriker.1992,Katz.1992}: $\rhoSFR = \epsff\, \rho/\tff$, where $\tff$ is the freefall time at $\rho$. Thus, we choose $\beta=1.5$ for our fiducial simulations but also explore how our results depend on $\beta$ by varying it in the range $\beta=1 \div 2.5$. We adopt a fiducial value of $\epsff=1\%$ motivated by the estimates in observed star-forming molecular clouds \citep{Evans.etal.2009,Evans.etal.2014,Heiderman.etal.2010,Lada.etal.2010,Lee.etal.2016,Leroy.etal.2017,Utomo.etal.2018}. However, we also consider the results of simulations with a much higher value of $\epsff=100\%$. 

\begin{figure*}
\centering
\includegraphics[width=0.56\textwidth]{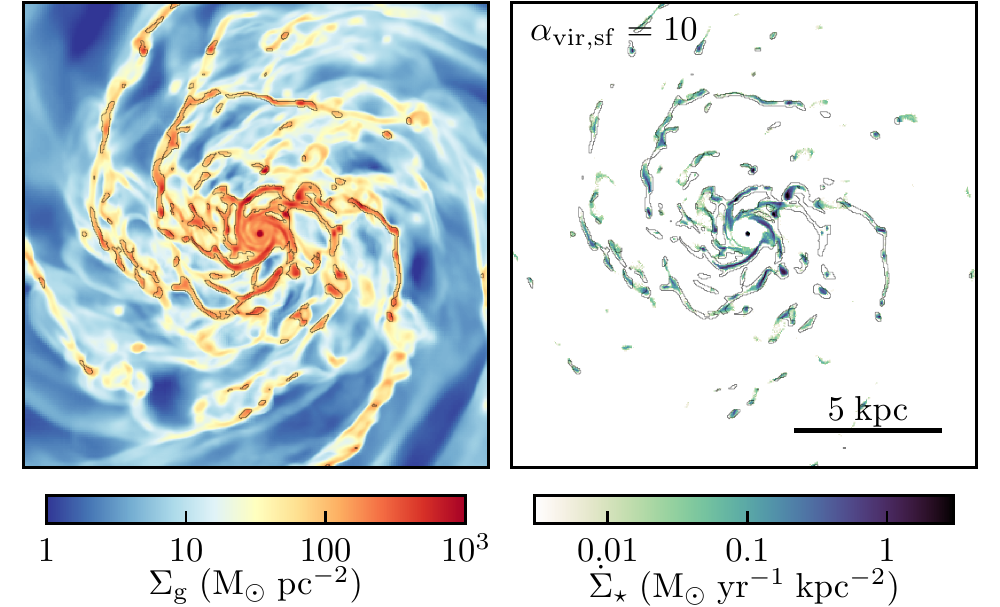}%
\includegraphics[width=0.43\textwidth]{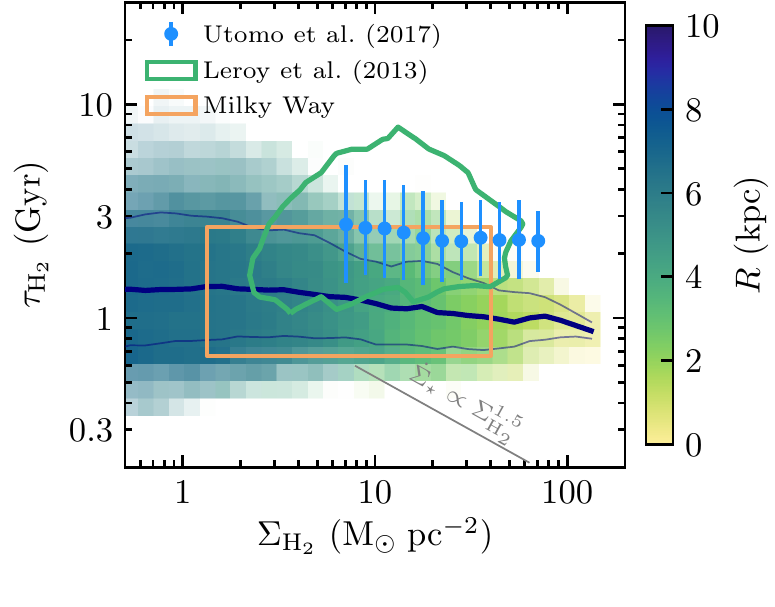}%
\caption{\label{fig:ksr-avir10} Results of our fiducial simulation with the star formation threshold in virial parameter, $\avir < \avirsf = 10$. The depletion time of molecular gas is almost independent of $\SH2$ and thus the molecular KSR is almost linear in this simulation. The two face-on maps on the left show the surface densities of total gas, $\Sg$, and SFR, $\SSFR$. SFR is averaged over $10\Myr$ as explained in Section~\ref{sec:sims}. The surface density of molecular gas, $\SH2$, is computed using the \citet{KMT1} model (see Section~\ref{sec:sims}), and the gray contours indicate regions where $\SH2/\Sg > 30\%$. The right panel shows the depletion time of molecular gas, $\tH2 = \SH2/\SSFR$, averaged on $1\kpc$ scale as a function of $\SH2$. The distribution is averaged over 11 snapshots between 500 and $600\Myr$ and is colored according to the average galactocentric radius $R$. Dark blue lines show the running median (thick line) and 16$^{\rm th}$ and 84$^{\rm th}$ percentiles (thin lines). Blue points show the running median in the EDGE-CALIFA sample of galaxies from \citet{Utomo.etal.2017}. The green contour shows 68\% of points from \citet{Leroy.etal.2013}; the median of this sample is very similar to \citet{Utomo.etal.2017}, and therefore, it is not shown here. The orange rectangle shows $\tH2 \sim 0.7\textrm{--}2.7\Gyr$ estimated for the Milky Way using $\SFR \sim 1\Msunyr$ and $\MH2 \sim 10^{9} \Msun$ \citep[before correction for helium;][]{Heyer.Dame.2015} and adopting a factor of 2 uncertainty in $\tH2$ \citep[Figure~7 in][]{Kennicutt.Evans.2012}. }
\end{figure*}

Feedback from young stars is modeled by injection of momentum and thermal energy in the amounts calibrated against simulations of supernova remnant evolution in a non-uniform medium by \citet{Martizzi.etal.2015}. To compensate for the numerical loss of momentum due to advection errors\footnote{We found that in idealized simulations of a stellar particle moving through a uniform medium, the loss of momentum can be as large as a factor of $\sim3\textrm{--}10$, especially if the particle moves diagonally to the grid.} and for additional effects, such as momentum boost due to supernova clustering \citep{Gentry.etal.2017,Gentry.etal.2018} and cosmic rays \citep{Diesing.Caprioli.2018}, we multiply the momentum predicted by \citet{Martizzi.etal.2015} by a fiducial factor of 5.  To estimate the effect of feedback strength on the molecular KSR slope, we also ran simulations with a 5 times stronger feedback (i.e., the \citealt{Martizzi.etal.2015} momentum multiplied by 25) and simulations without feedback. The resulting radial momentum is distributed among all immediate neighbors\footnote{For a uniform grid, the number of neighbors receiving momentum is 26, but it can be larger depending on the local refinement structure of the grid.} of the cell hosting the supernova. 
In addition to supernova feedback, we also account for the mass loss by stellar particles following \citet{Leitner.Kravtsov.2011} and inject mass and the corresponding momentum resulting from the motion of the stellar particle with respect to the gas into the cell hosting the stellar particle.

To obtain the $\SSFR$--$\SH2$ relation in our simulations, we measure the SFR surface density from the distribution of stars younger than $10\Myr$. This time interval is chosen to approximate the temporal averaging of SFR used in the extragalactic observations to which we will compare our results.\footnote{Specifically, 10$\Myr$ corresponds to the H$\alpha$ star formation indicator \citep[see, e.g., Table 1 in][]{Kennicutt.Evans.2012}. Extragalactic observations usually adopt a combination of H$\alpha$ and infrared indicators, which corresponds to the SFR averaged over a somewhat longer time interval. However, we checked that our results remain almost unchanged when we average SFRs over 30$\Myr$ instead of 10$\Myr$.} Molecular gas surface density is measured by integrating molecular density in all cells along the direction perpendicular to the disk. The molecular density in each cell is defined as $\rho_{\rm H_2} = f_{\rm H_2} \rho$, where $\rho$ is the local total density of gas, and $f_{\rm H_2}$ is computed using the model of \citet{KMT1,KMT2} and \citet{McKee.Krumholz.2010}: $f_{\rm H_2} = \max[0,(1-0.75s/(1+0.25s))]$ with $s\approx1.8/\tau_{\rm c}$ and $\tau_{\rm c} = 320(\rho\Delta/{\rm g\;cm^{-2}})$ at solar metallicity. Although such a definition of molecular gas density includes the contribution from helium, for simplicity, we will denote quantities related to molecular gas with a subscript ``H$_2$.''

In Section~\ref{sec:ksr}, we will connect the origin of the linear molecular KSR to gas evolution between different states in the ISM. To explore this connection, we measure relevant timescales of gas evolution in these states using passive gas tracers, which are initialized proportionally to gas density after $300\Myr$ of disk evolution and then stochastically exchanged between adjacent cells with the probability proportional to mass flux at each time step \citep{Genel.etal.2013}. In particular, for each tracer, we measure the duration of molecular stages as
\begin{equation}
\tmH2 = \int \limits_{t_1}^{t_2} \fH2(t)\, dt,
\end{equation}
where $\fH2$ is the molecular mass fraction of the cell hosting the gas tracer at a given time step, and $t_1$ and $t_2$ are the subsequent moments when the gas tracer crosses the $n=10\cc$ threshold\footnote{For our choice of parameters, the \citet{KMT1} model predicts that the molecular mass fraction is 0 for $n<10\cc$ and it sharply increases for $n>10\cc$}.
Next, we define the total time in the star-forming state during $\tmH2$ as
\begin{equation}
\tsf = \int \limits_{t_1}^{t_2} \Theta_{\rm sf}(t)\, dt,
\end{equation}
where at each time step $\Theta_{\rm sf} = 1$ if a gas tracer resides in a cell that satisfies the star formation criterion, and $\Theta_{\rm sf} = 0$ otherwise. Finally, we also accumulate integral star formation efficiency at each molecular stage as
\begin{equation}
\epsint = \int \limits_{t_1}^{t_2} \frac{\Theta_{\rm sf}(t)}{\tdep(t)}\, dt,
\end{equation}
where $\tdep = \rho/\rhoSFR$ is the local depletion time in star-forming gas; $\tdep = \tff/\epsff$ for our fiducial $\beta=1.5$ in Equation~(\ref{eq:rhoSFR-slope}).

%--------------------------------------------------------------------
\section{Molecular KSR in simulations}
\label{sec:results}
%--------------------------------------------------------------------

\begin{figure*}
\centering
\includegraphics[width=0.378\textwidth]{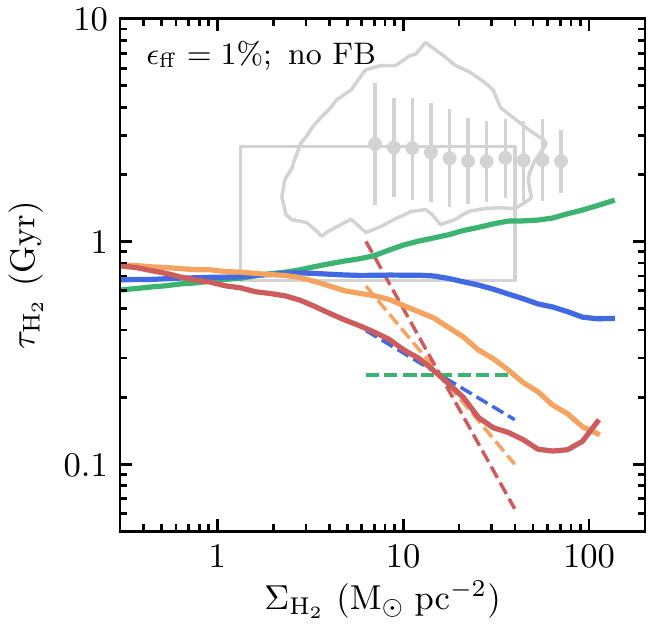}%
\includegraphics[width=0.311\textwidth]{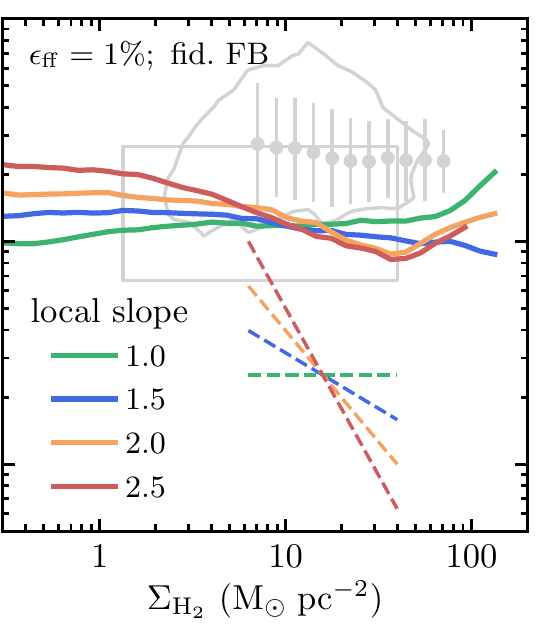}%
\includegraphics[width=0.311\textwidth]{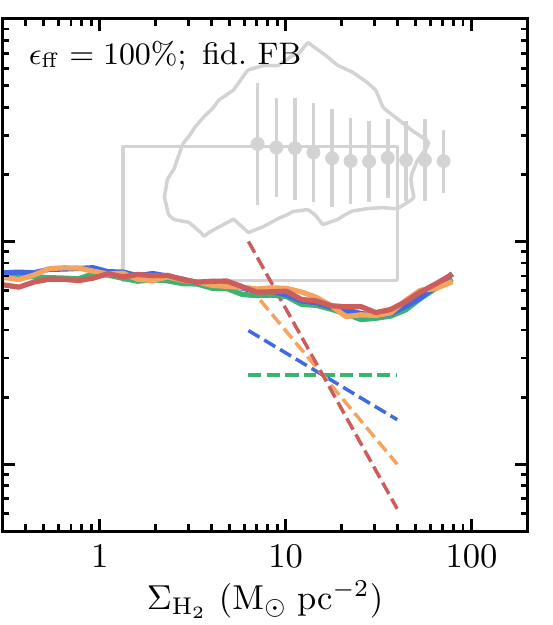}
\caption{\label{fig:slopes} Dependence of the molecular KSR in simulations with $\avirsf=10$ threshold on the slope of the local star formation relation: $\rhoSFR \propto \rho^{\beta}$, where $\beta$ is varied from 1.0 to 2.5 (see Equation~\ref{eq:rhoSFR-slope}). The adopted slopes are indicated by dashed lines, which show the scaling of the local depletion time, $\rho/\rhoSFR \propto \rho^{1-\beta}$. Three panels show simulations with different $\epsff$ values and feedback strengths per supernova, with the total feedback budget per local freefall time increasing from left to right. In simulations without feedback (left panel), the median $\tH2(\SH2)$ on kiloparsec scales (thick lines) depends on the local slope. As the feedback budget increases, the sensitivity to the local slope becomes weaker in our fiducial simulation (middle panel) and completely disappears in simulations with 100 times larger local $\epsff$ (right panel). Remarkably, in the latter regime of very efficient feedback, $\tH2$ becomes almost independent of $\SH2$ as observed in real galaxies (gray contours and error bars; see the legend in the right panel of Figure~\ref{fig:ksr-avir10}).}
\end{figure*}

Figure~\ref{fig:ksr-avir10} shows the results of our fiducial simulation with the star formation threshold in gas virial parameter, $\avir < \avirsf = 10$, $\epsff=1\%$, and $\beta=1.5$. The left two panels show the surface densities of total gas, $\Sg$, and SFR, $\SSFR$. The surface density of molecular gas, $\SH2$, reflects the distribution of dense gas and, for reference, the gray contours indicate the regions where the molecular gas fraction is larger than $30\%$. The right panel shows the depletion time of molecular gas, $\tH2 = \SH2/\SSFR$, as a function of $\SH2$ where both $\SSFR$ and $\SH2$ are averaged on $1\kpc$ scale using a 2D Gaussian filter with a width of $1\kpc$.

The figure shows that $\tH2$ in our fiducial simulation is almost independent of $\SH2$ and thus the molecular KSR is almost linear, $\SSFR \propto \SH2$, in agreement with the molecular KSR of observed galaxies at similar $\SH2$. This nearly linear slope is not imposed by the choice of our star formation prescription and its parameters. Instead, the near-linear slope emerges from a significantly steeper local star formation relation, $\rhoSFR = \epsff\, \rho/\tff \propto \rho^{1.5}$. If the molecular KSR reflected this small-scale slope, it would have the slope shown by the thin gray line in the right panel, which is clearly much steeper than the slope we measure.

To stress this point, in the middle panel of Figure~\ref{fig:slopes} we show that the $\SSFR$--$\SH2$ relation remains close to linear even when the local SFR is assumed to have a steeper dependence on gas density:  $\rhoSFR \propto \rho^2$ and $\propto \rho^{2.5}$. Such weak sensitivity of the $\SSFR$--$\SH2$ slope to its local value provides a counter-example to the argument that the KSR on $\sim$kpc scales simply reflects the local star formation relation \citep{Schaye.DallaVecchia.2008,Gnedin.etal.2014} and demonstrates that the KSR slope can be nearly independent of the slope of the local relation.

The left and right panels of Figure~\ref{fig:slopes} show that the emergence of the linear KSR slope and its independence of the local star formation relation depend on the efficiency of star formation and the corresponding efficacy of stellar feedback. Indeed, in simulations without feedback (left panel), the molecular KSR slope on kiloparsec scale becomes quite sensitive to the assumed local slope $\beta$. On the other hand, in simulations with $\epsff=100\%$ where feedback is very efficient (right panel), kiloparsec-scale molecular KSR becomes completely insensitive to the local star formation relation, and its slope stays close to linear. We have also checked that the relation remains linear when we increase feedback momentum input per supernova instead of increasing $\epsff$ (Appendix~\ref{app:sf-fb}) and when we continuously vary $\epsff$ with $\avir$ instead of using a sharp threshold (Appendix~\ref{app:turb}). We also find that the slope remains linear as long as the molecular KSR is averaged on $\gtrsim500\pc$ scales (see Appendix~\ref{app:scale}).

\begin{figure*}
\centering
\includegraphics[width=0.56\textwidth]{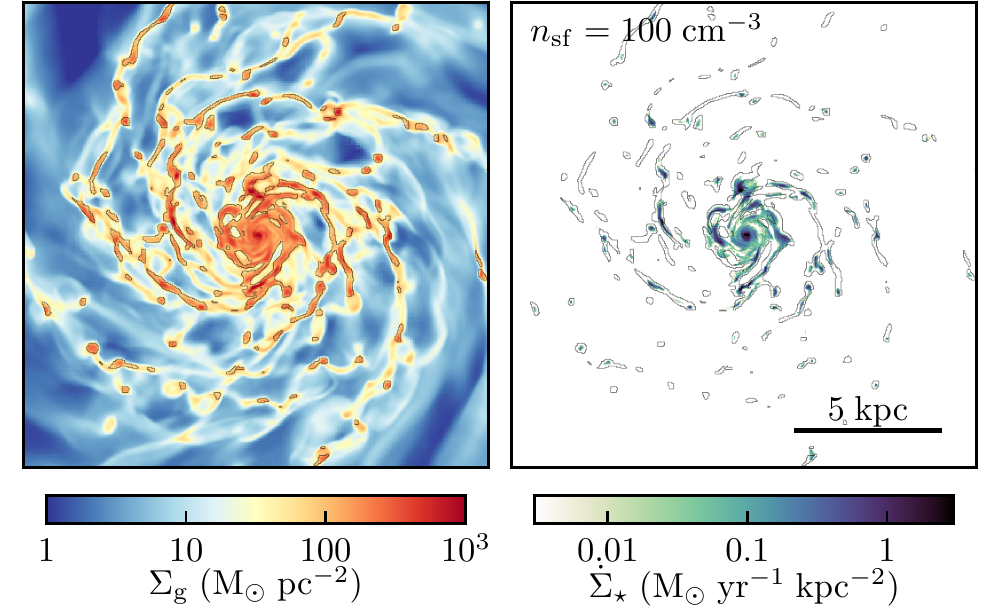}%
\includegraphics[width=0.43\textwidth]{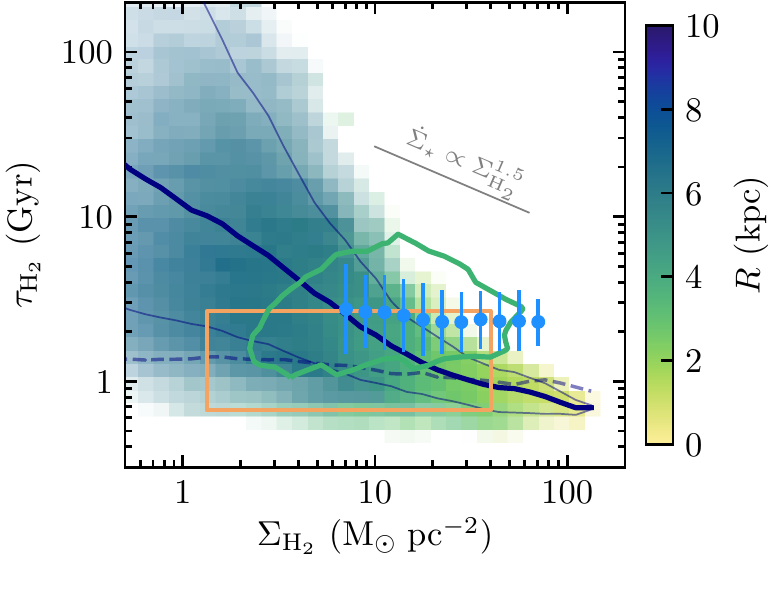}%
\caption{\label{fig:ksr-n100} Same as Figure~\ref{fig:ksr-avir10} but for a star formation threshold in gas density: $n > \nsf = 100\cc$. The dashed line in the right panel shows the median $\tH2$ from Figure~\ref{fig:ksr-avir10}. For a density threshold, the molecular KSR becomes significantly steeper than in the simulation with an $\avir$-based threshold. }
\end{figure*}

Apart from feedback, the choice of star formation threshold also plays a crucial role in shaping the kiloparsec-scale molecular KSR. For example, Figure~\ref{fig:ksr-n100} shows that when we choose a threshold in density, $n > \nsf = 100\cc$, instead of the $\avir$, the molecular KSR significantly steepens to $\SSFR \propto \SH2^{1.5}$ at $\SH2 > 10\Msunpc2$ (galactocentric radii of $R < 6\kpc$) and to even steeper slope at $\SH2 < 10\Msunpc2$ ($R > 6\kpc$). A similar steep molecular KSR was also found by \citet{Capelo.etal.2018}, who also used the $\nsf=100\cc$ star formation threshold. In addition, we find that the slope steepens for larger values of $\epsff$ and larger values of the $\nsf$ threshold (see Appendix~\ref{app:sf-fb}). In other words, the emergence of the linear KSR and its insensitivity to local star formation parameters occur only when we use the $\avir$-based threshold.

The qualitatively different behavior of the molecular KSR in simulations with $\avir$- and $\nsf$-based thresholds stems from different distributions of star-forming gas in simulations with different thresholds, which is evident from the $\SSFR$ maps in Figures~\ref{fig:ksr-avir10} and \ref{fig:ksr-n100}. In the simulation with the constant $\nsf$ threshold, the SFR is more centrally concentrated, and a significant fraction of molecular gas in the outskirts of the disk is not forming stars. This results in longer depletion times at large radii and steeper $\SSFR$--$\SH2$ relation.

The results presented above demonstrate that the slope of the molecular KSR has a nontrivial origin and is shaped by several factors. To understand the trends of the slope with the choice of star formation threshold, $\epsff$ value, and feedback strength, we will first examine our results from a different angle using the language of the PDF of gas properties (next Section). We will then discuss how our simulation results can be understood in the simple physical framework that we developed to explain the long gas depletion times of galaxies (Section~\ref{sec:ksr}).

%--------------------------------------------------------------------
\section{Connecting the kiloparsec-scale KSR with gas PDF and star formation relation on small scales}
\label{sec:pdf}
%--------------------------------------------------------------------

Before we proceed to discuss the physical interpretation of the simulation results presented above, we will examine how kiloparsec-scale KSR is connected to the local SFR and PDF of gas properties. This connection not only clearly illustrates the interdependencies of different quantities via explicit equations but also elucidates the conditions required for the linearity of the molecular KSR. To this end, we will consider the depletion time of gas in star-forming regions, $\taust=\Ssf/\SSFR$, and the mass fraction of molecular gas in these regions, $\fsf=\Ssf/\SH2$, as it is the ratio of these quantities that defines the global depletion time of molecular gas: 

\begin{equation}
\label{eq:tauH2_fsf}
\tH2 = \frac{\SH2}{\SSFR}=\frac{\SH2}{\Ssf}\,\frac{\Ssf}{\SSFR}=\frac{\taust}{\fsf}.
\end{equation}

The results presented in this section do not explain the simulation results presented above, but rather restate these results in a different useful way. Readers more interested in the physical interpretation of the results can proceed to Section~\ref{sec:ksr}. A detailed derivation of the equations used below is presented in the Appendix~\ref{app:pdf}.

%--------------------------------------------------------------
\subsection{Molecular Depletion Time and PDF of Gas Properties}
%--------------------------------------------------------------

To quantify the relation between molecular gas depletion time and gas PDF, we note that both $\fsf$ and $\taust$ in Equation~(\ref{eq:tauH2_fsf}) are the averages over the distribution of gas in the ISM patches. In particular, $\fsf$ can be expressed in terms of the PDF of molecular gas, $P_{\rm H_2}$:
\begin{equation}
\label{eq:fsf-PDF}
\fsf = \iint_0^\infty \Theta_{\rm sf}(\rho,\q) P_{\rm H_2}(\rho,\q|\SH2) d\rho d\q,
\end{equation}
where the star formation threshold, $\Theta_{\rm sf}(\rho,\q)$, can depend on the density, $\rho$, and any other local properties of the region, $\q$, such as temperature, turbulent velocity, etc. For our choice of thresholds, $\Theta_{\rm sf}(\rho,\sigma) = \theta(\avirsf-\avir(\rho,\sigma))$ and $\Theta_{\rm sf}(\rho) = \theta(\rho-\rhosf)$, where $\theta$ is the Heaviside step function: $\theta(x)=1$ for $x>0$ and $\theta(x)=0$ for $x<0$.\footnote{In general, $\Theta_{\rm sf}$ can also parameterize smooth transitions to the star-forming state by continuously changing between 0 and 1.} $P_{\rm H_2}(\rho,\q|\SH2)$ in the above equation is the mass-weighted PDF of the molecular gas averaged between patches with surface density $\SH2$; the PDF shape in general will depend on $\SH2$. 

Similarly, $\taust$ can be written as
\begin{equation}
\label{eq:taust-PDF}
\frac{1}{\taust} = \int_0^\infty \frac{1}{\tdep} P_{\rm sf}(\rho|\SH2) d\rho \propto \int_0^\infty \rho^{\beta-1} P_{\rm sf}(\rho|\SH2) d\rho,
\end{equation}
where $\tdep = \rho/\rhoSFR \propto \rho^{1-\beta}$ is the local depletion time,\footnote{The averaging of inverse $\tdep$ results from the averaging of local star formation rates $\rhoSFR=\rho/\tdep$: $\taust^{-1} = \SSFR/\Ssf = \int \tdep^{-1} \rho dV / \int \rho dV = \langle \tdep^{-1} \rangle_{\rm sf}$, where the integrals are taken over star-forming regions in an ISM patch.} and the mass-weighted density PDF of star-forming gas is defined as
\begin{equation}
\label{eq:sf-PDF}
P_{\rm sf}(\rho|\SH2) = \frac{\int \Theta_{\rm sf}(\rho,\q)P_{\rm H_2}(\rho,\q|\SH2) d\q}{\iint_0^\infty \Theta_{\rm sf}(\rho,\q)P_{\rm H_2}(\rho,\q|\SH2) d\rho d\q}.
\end{equation}

First, note that the dependencies of $\fsf$ and $\taust$ on $\SH2$ in Equations~(\ref{eq:fsf-PDF}) and (\ref{eq:taust-PDF}) are rather nontrivial. To obtain a linear KSR, these dependencies must be similar and cancel out in Equation~(\ref{eq:tauH2_fsf}), which means that the gas PDF must scale with $\SH2$ in a very particular way. Second, the observed independence from the local slope $\beta$ in simulations with efficient feedback implies that the trends of $\fsf$ and $\taust$ must remain similar when $\beta$ is changed. This effect is also nontrivial because $\taust$ depends on $\beta$ explicitly (Equation~\ref{eq:taust-PDF}), while $\fsf$ depends on $\beta$ only implicitly via the effect of $\beta$ on the gas PDF. Finally, Equations (\ref{eq:fsf-PDF}--\ref{eq:sf-PDF}) show that $\fsf$ and $\taust$ explicitly depend on the shape of the star formation threshold, $\Theta_{\rm sf}$. This means that, given a specific gas PDF, the above cancellation can occur for some thresholds but not for the others. In the next section, we will examine the trends of the gas PDF, $\taust$, $\fsf$, and $\tH2$ observed in our simulations.

%--------------------------------------------------------------------
\subsection{Trends of the Gas PDF, $\taust$, $\fsf$, and $\tH2$ in Simulations}
\label{sec:sims-pdf}
%--------------------------------------------------------------------

\begin{figure*}
\centering
\includegraphics[width=\textwidth]{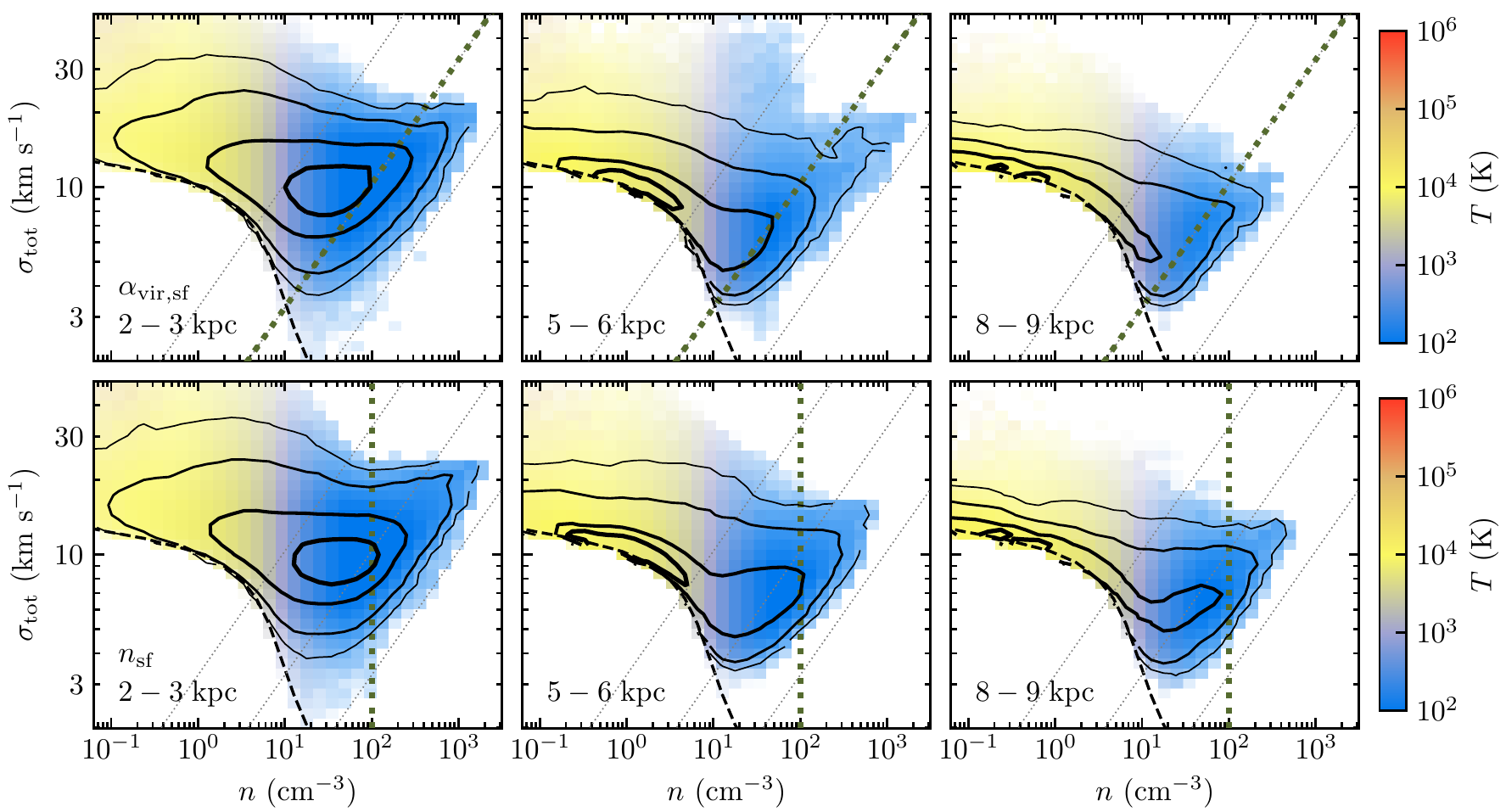}
\caption{\label{fig:phases} Dependence of gas distribution on galactocentric radius (and therefore on the average gas density) in the simulations with $\avirsf=10$ (top row) and $\nsf=100\cc$ star formation threshold (bottom row). Each column of panels shows the distribution averaged between 500 and $700\Myr$ in $1\kpc$ wide annuli at different galactocentric radii: $R = 2\textrm{--}3\kpc$ (left), $5\textrm{--}6\kpc$ (middle), and $8\textrm{--}9\kpc$ (right). Contours show 25\%, 68\%, 95\%, and 99\% of the PDF. The color of the distribution indicates the average temperature, and the blue region (cold gas) roughly indicates the distribution of molecular gas. The dashed line close to the lower envelope of the distribution shows the median temperature as a function of density. The thick dotted line shows the star formation threshold adopted in these simulations; star-forming gas resides to the right of the threshold. Thin dotted lines indicate constant values of the virial parameter: $\avir=100$, 10, and 1 from left to right. }
\end{figure*}  

The dependence of the gas PDF, $\taust$, $\fsf$, and $\tH2$ on $\SH2$, the local star formation slope $\beta$, and the star formation threshold can be explored in simulations directly, as all of the relevant quantities can be measured. In particular, we will consider these trends in two simulations from our suite: the fiducial run in which the KSR is linear (Figure~\ref{fig:ksr-avir10}) and the run with the same parameters but with the density-based star formation threshold in which the molecular KSR is steep (Figure~\ref{fig:ksr-n100}). 

Figure~\ref{fig:phases} shows the distribution of gas in these simulations in the plane of gas density and total velocity dispersion, $\stot = \sqrt{\cs^2+\st^2}$, which includes both thermal and subgrid turbulent velocities. The distributions are averaged in $1\kpc$-wide annuli at different galactocentric radii, $R$, and three horizontal panels show three representative radii. Given that the gas surface density exponentially decreases with increasing $R$, these distributions correspond to widely different $\SH2$ values.

\begin{figure}
\centering
\includegraphics[width=\columnwidth]{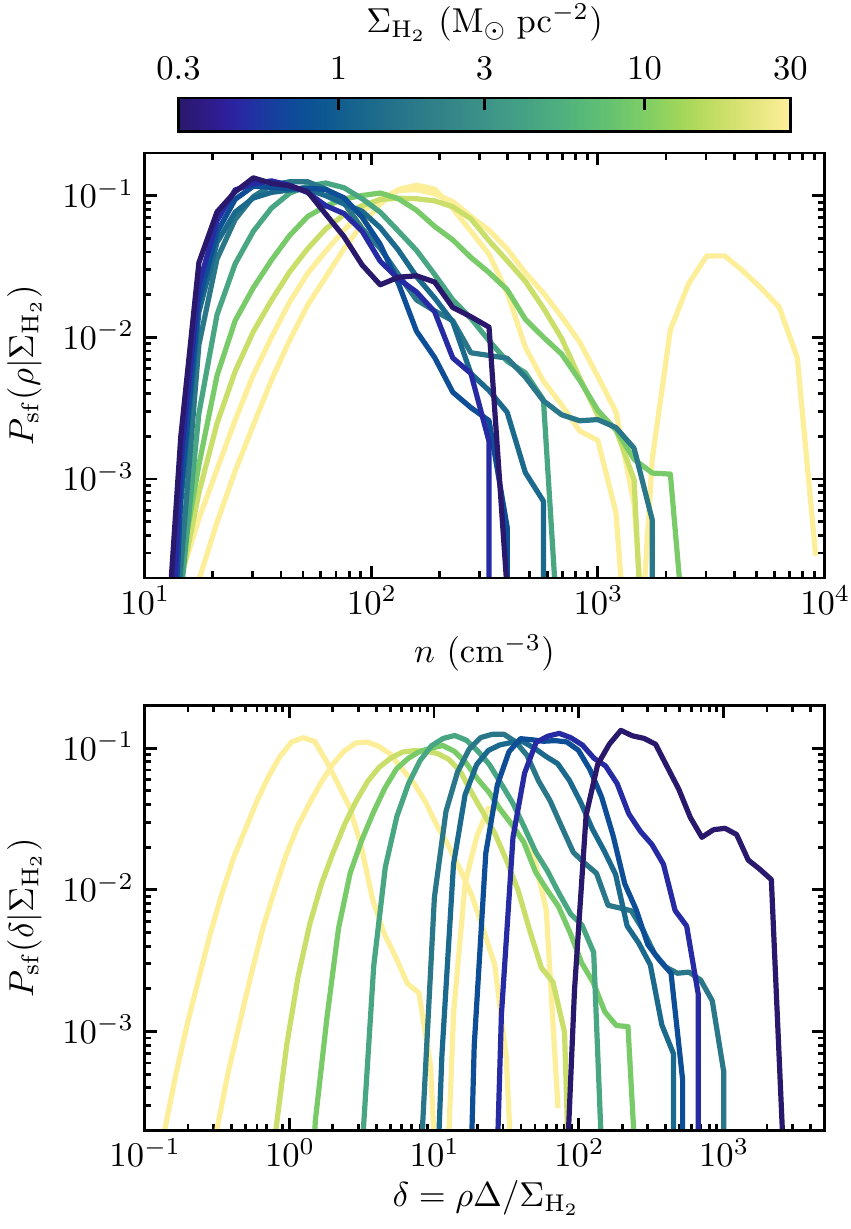}
\caption{\label{fig:avir_pdfs} Density PDFs of star-forming gas (Equation~\ref{eq:sf-PDF}) in concentric $1\kpc$ wide annuli in the simulation with the $\avirsf$ threshold. PDFs are averaged between 500 and $700\Myr$ and colored according to the average density of molecular gas in each annulus, $\SH2$. The top panel shows that PDFs are not fixed but scale with $\SH2$, while the bottom panel demonstrates that this scaling is weaker than self-similar. If the PDF scaled self-similarly, its shape would be fixed as a function of $\rho/\SH2$. To make this ratio dimensionless, we multiply it by our resolution $\Delta = 40\pc$. The PDF in the central region ($R<1\kpc$, pale yellow) is bimodal, with a prominent bump at $n\sim(2-7)\times10^3\cc$, which corresponds to a central concentration of high-$\st$ gas that does not form stars according to the $\avir<10$ criterion. When we use a density threshold (Figure~\ref{fig:n_pdfs}), such a bump does not form because all gas at such densities would rapidly form stars.}
\end{figure}

As the top row in the figure shows, in our fiducial simulation, the gas PDF does change with changing $\SH2$, which results in trends of $\taust$ and $\fsf$ with $\SH2$. In particular, according to Equation~(\ref{eq:taust-PDF}), the trend of $\taust$ is due to the scaling of the star-forming part of the PDF. To explore the $\taust$ trend in more detail, in Figure~\ref{fig:avir_pdfs} we plot these parts of the PDF (integrated over $\stot$) in concentric $1\kpc$-wide annuli out to $R=10\kpc$. 

If star-forming gas PDF were independent of $\SH2$, $P_{\rm sf}(\rho|\SH2) = F(\rho)$, Equation~(\ref{eq:taust-PDF}) implies that $\taust$ would also be independent of $\SH2$. This case corresponds to the ``counting argument'' often used as an explanation for the linearity of the molecular KSR (see Section~\ref{sec:intro}). However, the top panel of Figure~\ref{fig:avir_pdfs} shows that the PDF is not fixed but shifts to higher density with increasing $\SH2$. Therefore, $\taust$ is not constant but decreases with $\SH2$.

On the other hand, if the PDF scaled self-similarly, $P_{\rm sf}(\rho|\SH2) = F(\rho/\SH2)/\SH2$ where $F$ is a function of a fixed shape and peak location, then $\taust$ would inherit the slope from the local star formation relation: $\taust \propto \SH2^{1-\beta} = \SH2^{-0.5}$ for $\beta=1.5$, as is explicitly shown in Appendix \ref{app:pdf} \cite[see also][]{Gnedin.etal.2014}. However, the bottom panel of Figure~\ref{fig:avir_pdfs} shows that PDFs for different $\rho/\SH2$ do not coincide, with their peak changing significantly with $\SH2$. This strong deviation from self-similar scaling results in a scaling of $\taust$ that is weaker than the self-similar expectation of $\taust \propto \SH2^{-0.5}$.

We find that in our fiducial simulation, $\taust$ scales as $\taust \propto \SH2^{-0.2}$. If the star-forming fraction $\fsf$ were independent of $\SH2$, this scaling would be enough to make the KSR noticeably nonlinear: $\SSFR \propto \SH2^{1.2}$. However, $\fsf$ scales with $\SH2$ similarly to $\taust$, as can be seen in Figure 9 in \citet{Semenov.etal.2017}. Given that $\tH2=\taust/\fsf$ (Equation~\ref{eq:tauH2_fsf}), these scalings cancel, resulting in $\tH2$ almost independent of $\SH2$ and a nearly linear molecular KSR.

\begin{figure}
\centering
\includegraphics[width=\columnwidth]{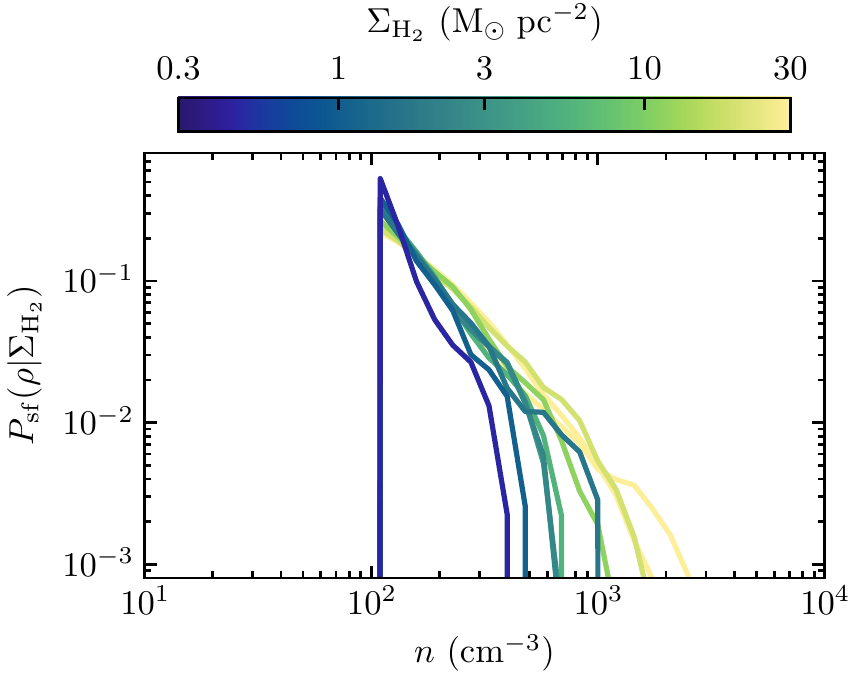}
\caption{\label{fig:n_pdfs} Same as the top panel in Figure~\ref{fig:avir_pdfs} but for the simulation with the $\nsf=100\cc$ star formation threshold. The dependence of PDF on $\SH2$ is much weaker compared to the simulation with the $\avirsf=10$ threshold. }
\end{figure}

We find that in the simulations with an $\avir$-based threshold, such cancellation holds for all values of the slope of the local star formation relation $\beta$ as long as feedback is as strong as in the fiducial simulation or stronger (see Section~\ref{sec:results} and Appendix~\ref{app:sf-fb}). Such cancellation, however, does not happen in the simulations without feedback, in which the slope of the molecular KSR becomes sensitive to the local slope $\beta$, and in all of the simulations with the density-based star formation threshold. In the latter case, the density threshold defines only the high-density tail of distribution as star-forming gas (see the bottom row of Figure~\ref{fig:phases}). Therefore, the PDF of star-forming gas and $\taust$ become almost independent of $\SH2$, as shown in Figure~\ref{fig:n_pdfs}. However, the star-forming fraction of molecular gas does increase at higher $\SH2$, and therefore $\tH2=\taust/\fsf$ decreases with increasing $\SH2$, resulting in a superlinear molecular KSR. 

Results presented in this section clearly show that the linear slope of molecular KSR in the simulations with an $\avir$-based star formation threshold and efficient feedback is rather nontrivial and results from a cancellation of trends of physical properties controlling $\tH2$ with both $\beta$ and $\SH2$. The cancellation occurs only when feedback is efficient, and it depends on the choice of the star formation threshold.

The language of gas PDF is direct and clearly shows the relation between large-scale observables and small-scale properties of gas and star formation. However, we cannot use it to fully explain our simulation results because this requires knowledge of the exact functional form of the gas PDF, its dependence on $\SH2$, and its response to feedback. Therefore, in what follows, we will adopt an approximate approach that can qualitatively explain several of our key results.

%--------------------------------------------------------------------
\section{The physics of the molecular KSR slope}
\label{sec:ksr}
%--------------------------------------------------------------------

In this section, we show that several key results of our simulations  can be understood using the theoretical framework that was introduced in \citet{Semenov.etal.2017,Semenov.etal.2018} to explain long depletion times of gas in observed and simulated galaxies. The model is based on the conservation of mass and considers dynamic gas cycling between star-forming and non-star-forming states. In this model, the depletion time of total gas in an ISM patch is $\tglob \sim \Nc (\tnsf + \tsf)$, where $\tnsf$ and $\tsf$ are the average times in non-star-forming and star-forming stages in each cycle, and the number of cycles, $\Nc \sim \taust/\tsf$, is set by the condition that after many cycles, gas must spend on average $\taust\equiv \Ssf/\SSFR$ in the star-forming state. The fraction of the initial gas parcel mass that is converted into stars during each cycle is correspondingly $\epsint\sim \tsf/\taust$. 

If the duration of star-forming stage on each cycle is short (e.g., due to efficient feedback) and local star formation in this state is inherently inefficient (i.e. $\taust$ is long), a typical gas parcel must go through a large number of such cycles ($\Nc\gg 1$) before it is incorporated into a star. This explains why the global gas depletion time is much longer than both the timescales of the ISM processes, $\tsf$ and $\tnsf$, and the local depletion time $\taust$. 

%--------------------------------------------------------------------
\subsection{Molecular KSR and Gas Evolution Timescales}
\label{sec:cycling:eqs}
%--------------------------------------------------------------------

In the dynamical framework outlined above, the star-forming mass fraction of molecular gas can be expressed as the relative time in star-forming and molecular stages on each cycle, $\tsf$ and $\tmH2$:
\begin{equation}
\label{eq:fsf_tH2}
\fsf \equiv \frac{\Ssf}{\SH2} \sim \frac{\tsf}{\tmH2}.
\end{equation}
Here, $\tmH2$ is the duration of the molecular stage during one evolution cycle---i.e., the time between the moment when gas becomes molecular and the moment when it becomes atomic again. The time $\tsf$ is the total time during one cycle that gas spends in the star-forming state. Note that during $\tmH2$ a given gas parcel may remain non-star-forming ($\tsf=0$) or become star-forming one or multiple times. In the latter case, $\tsf$ is the sum of all star-forming stages that a gas parcel experienced. 

Given these definitions, Equation~(\ref{eq:tauH2_fsf}) can be rewritten as 
\begin{equation}
\label{eq:tauH2_tH2}
\tH2 = \frac{\taust}{\fsf} \sim \frac{\taust}{\tsf}\, \tmH2\sim \frac{\tmH2}{\epsint}.
\end{equation}

To express $\tsf$ via the timescales of the processes driving gas evolution, we note that the amount of star-forming gas in ISM patches, $\Ssf$, is regulated by the combined effect of gas consumption at the rate $\SSFR$ and dispersal of star-forming gas by feedback and dynamical processes (e.g., passages of spiral arms) at the rates of $\Fmfb$ and $\Fmdyn$, respectively. Following \citet{Semenov.etal.2017,Semenov.etal.2018}, we parameterize these as 
\begin{equation}
\label{eq:Fmfb} \Fmfb = \xi \SSFR = \xi\, \frac{\Ssf}{\taust}, \qquad \Fmdyn = \frac{\Ssf}{\tmd},
\end{equation}
where $\xi$ is the ``mass-loading factor'' of star-forming gas dispersal by feedback, while $\tmd$ is the characteristic timescale of the dynamical dispersal of star-forming regions. The total rate of gas removal from star-forming regions is thus $\SSFR+\Fmfb+\Fmdyn$ and the characteristic
time that gas spends in the star-forming state can thus be expressed as
\begin{equation}
\label{eq:tsf}
\tsf \sim \frac{\Ssf}{\SSFR+\Fmfb+\Fmdyn} = \taust\, \left( 1 + \xi + \frac{\taust}{\tmd} \right)^{-1}.
\end{equation}

To relate these quantities to the depletion time of molecular gas, consider the characteristic time that gas spends in molecular form, $\tmH2 = \tmolnsf + \tsf$, where $\tmolnsf$ is the time that molecular gas spends outside of star-forming regions. Equations~(\ref{eq:tauH2_tH2}) and (\ref{eq:tsf}) then give
\begin{equation}
\label{eq:tauH2_sum}
\tH2 = (1+\xi)\, \tmolnsf  + \left( 1 + \frac{\tmolnsf}{\tmd} \right) \taust.
\end{equation}

This expression is analogous to Equation~(14) in \citet{Semenov.etal.2018} and has similar behavior. In particular, only the second term explicitly depends on the slope of the local star formation relation $\beta$ via the dependence of $\taust$ on $\beta$ (Equation~\ref{eq:taust-PDF}). The first term can depend on $\beta$ only implicitly. As we will detail below, the regimes in which one of these two terms dominates correspond to the regimes where molecular KSR is sensitive or insensitive to $\beta$ (Figure~\ref{fig:slopes}). As we will also show, Equation~(\ref{eq:tauH2_sum}) can help us understand why the molecular KSR is close to linear when an $\avir$-based star formation threshold is used and why it steepens for a density-based threshold.

%--------------------------------------------------------------------
\subsection{Dependence of Molecular KSR on the Local Slope $\beta$}
\label{sec:cycling:fb}
%--------------------------------------------------------------------

When feedback is strong ($\xi$ is large) and star formation is locally efficient, i.e. $\taust$ is short (due to, e.g., large $\epsff$), the first term in Equation~(\ref{eq:tauH2_sum}) dominates. Given that only $\taust$ in the second term explicitly depends on $\beta$ (Equation~\ref{eq:taust-PDF}), in this regime, $\tH2$ can be expected to only weakly depend on the local slope. Indeed, we showed that the sensitivity to $\beta$ disappears completely for the $\epsff=100\%$ compared to a weak dependence in the $\epsff=1\%$ runs (Figure~\ref{fig:slopes}). This insensitivity indicates that in the context of our framework, the first term does not depend on $\beta$ even indirectly. This can be understood as follows.  

\begin{figure}
\centering
\includegraphics[width=\columnwidth]{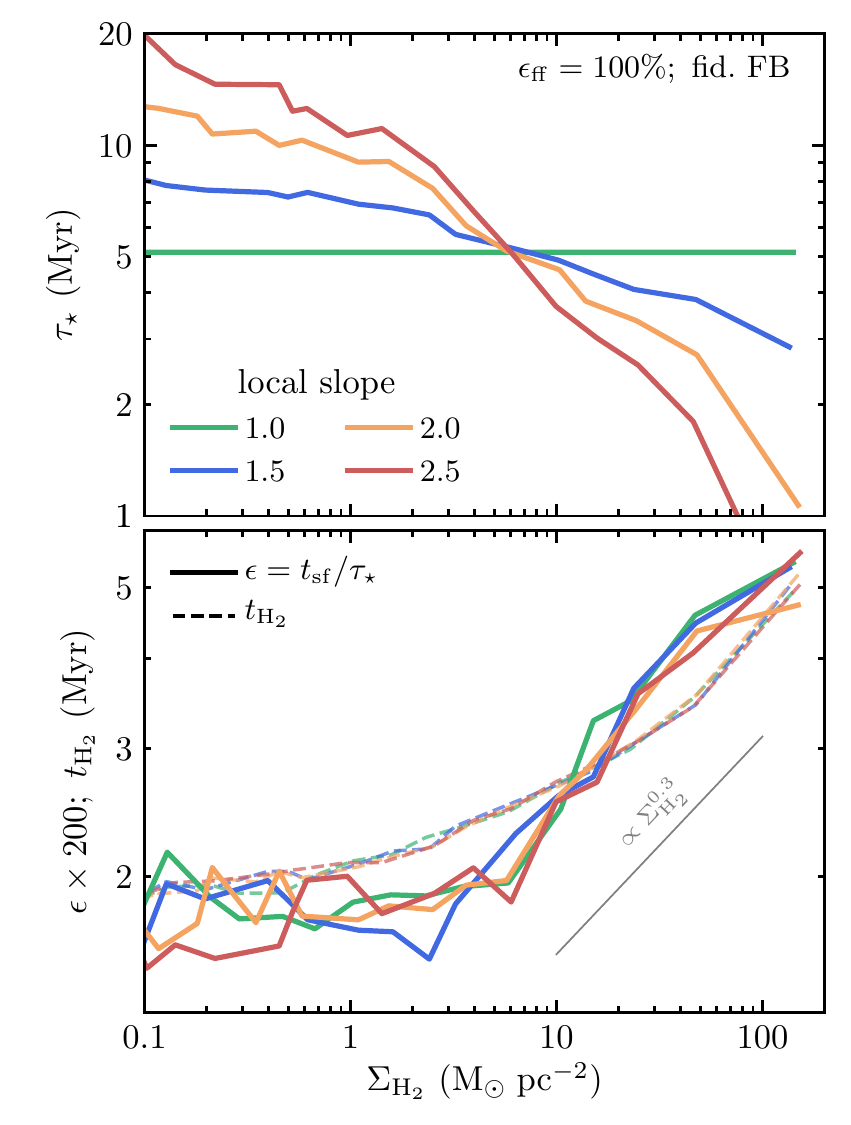}
\caption{\label{fig:taust-eps} Trends of $\taust$ (top panel), $\epsint = \tsf/\taust$ and $\tmH2$ (bottom panel) in our simulations with $\epsff=100\%$ and different slopes of the local star formation relation $\rhoSFR \propto \rho^\beta$, $\beta = 1$, 1.5, 2, and 2.5 (see Equation~\ref{eq:rhoSFR-slope}). We measure $\epsint$, $\tsf$, and $\tmH2$ for each molecular stage using gas-tracer particles as explained at the end of Section~\ref{sec:sims} and show their values averaged in concentric annuli. The stages are accumulated between 450 and 800 Myr of disk evolution. The figure illustrates two types of cancellation that result in a near-linear KSR. First, the variation of the $\taust$ trend with $\beta$ cancels out by $\tsf$ in $\epsint = \tsf/\taust$. Second, the trends of $\epsint$ and $\tmH2$ with $\SH2$ nearly cancel in $\tH2 \sim \tmH2/\epsint$, resulting in a nearly linear molecular KSR. }
\end{figure}

Equation~(\ref{eq:tsf}) shows that in this regime ($\xi\gg \taust/\tmd$), the characteristic time a gas parcel spends in a star-forming region in a single cycle is $\tsf\sim \taust/(1+\xi)\sim \taust\epsilon$ and corresponds to the time it takes to form enough young stars to disperse a typical star-forming region. For the higher SFR of larger $\beta$ values, and correspondingly smaller $\taust$, the young star mass fraction $\epsilon$ required for dispersal will be reached faster, and $\tsf$ will be shorter. Thus, both $\taust$ and $\tsf$ depend on $\beta$ in a similar way, so that this dependence cancels in $\epsint\sim \tsf/\taust$ rendering the integral stellar mass fraction required for dispersal roughly constant. Given that the lifetime of gas in the molecular state is not related to $\beta$, this cancellation is the main reason why $\tH2 \propto \tmH2\,\taust/\tsf$ is independent of $\beta$. The bottom panel of Figure~\ref{fig:taust-eps} demonstrates that $\epsint$ and $\tmH2$ are indeed independent of $\beta$ in the simulation with $\epsff=100\%$, even though $\taust$ does depend on $\beta$, as shown in the top panel.

\begin{figure}
\centering
\includegraphics[width=\columnwidth]{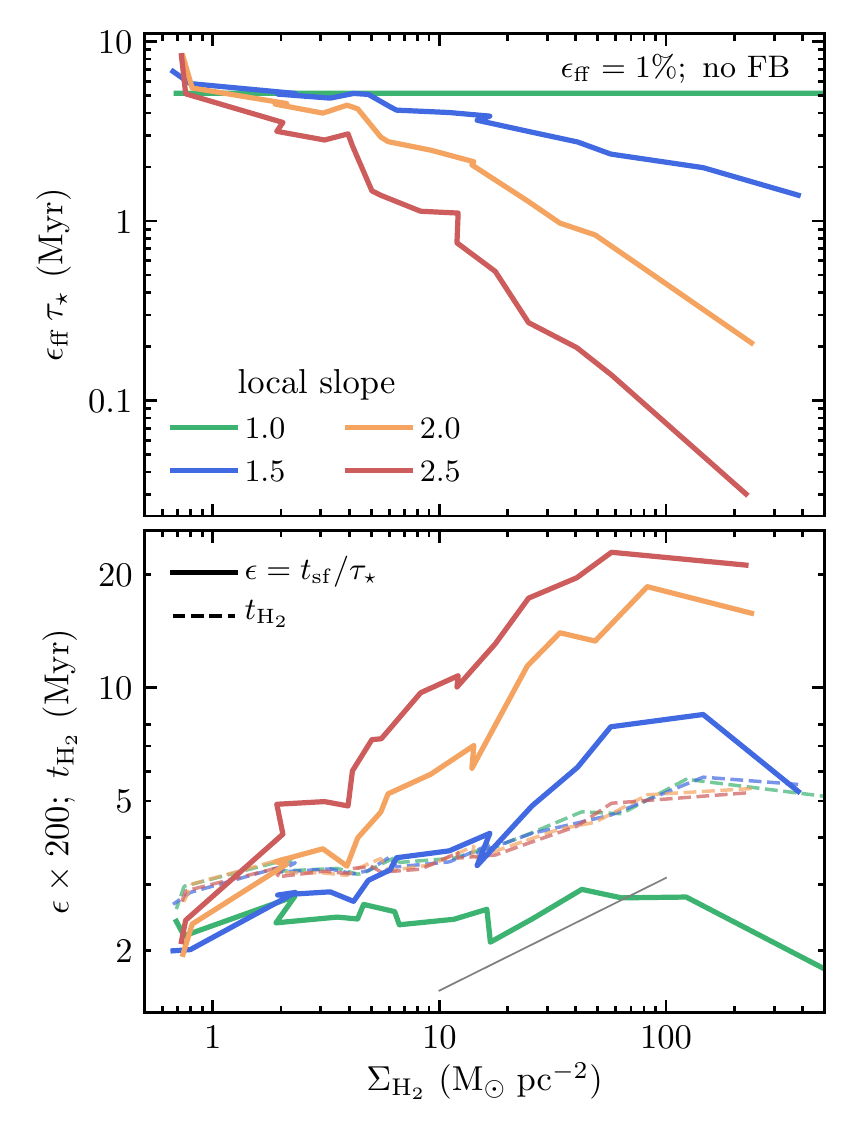}
\caption{\label{fig:taust-eps-nofb} Same as Figure~\ref{fig:taust-eps} but for simulations with $\epsff=1\%$ and no feedback. While the trends of $\tmH2(\SH2)$ remain almost the same as in Figure~\ref{fig:taust-eps}, there is no longer feedback-imposed cancellation of the trends in $\epsint = \tsf/\taust$, and $\epsint(\SH2)$ becomes strongly dependent on $\beta$, which leads to the dependence of $\tH2$ and the KSR slope on $\beta$ in simulations without feedback (left panel of Figure~\ref{fig:slopes}). To produce this plot, we accumulated molecular and star-forming stages of gas-tracers evolution over a shorter period of time (between 450 and 600 Myr) because for large $\beta$ and no feedback, the total gas mass changes more rapidly due to the short global depletion times. The local depletion times in the top panel are normalized by $\epsff$ to simplify the comparison with the trends in Figure~\ref{fig:taust-eps}. The thin gray line in the bottom panel repeats that from Figure~\ref{fig:taust-eps}. }
\end{figure}

When feedback is weak (small $\xi$) or star formation is locally inefficient ($\taust$ is long), the second term in Equation~(\ref{eq:tauH2_sum}) dominates. This term  is proportional to $\taust$, which explicitly depends on the local slope $\beta$. As shown in Figure~\ref{fig:taust-eps-nofb}, this dependence of $\taust$ on $\beta$ also implies the dependence of the molecular KSR on $\beta$ because $\tmH2(\SH2)$ does not depend on $\beta$, while in this regime $\epsint(\SH2)$ does. Their ratio $\tH2\sim\tmH2/\epsint$, therefore, is no longer close to constant, but depends on $\SH2$ in a way sensitive to $\beta$, as observed in our simulations without feedback (left panel of Figure~\ref{fig:slopes}). It is worth noting that although there is a value of $\beta$ at which the large-scale slope becomes close to linear in these simulations, this is coincidental and holds only for $\beta \approx 1.5$. Physically, $\tsf$ in this regime is controlled not by feedback but by dynamical processes that disperse star-forming and molecular regions, such as turbulent shear, differential rotation, expansion behind spiral arms, etc.

In  the  intermediate regime, in which the two terms in Equation~(\ref{eq:tauH2_sum}) are comparable, we can expect an intermediate weak sensitivity of the KSR to $\beta$, as is indeed observed in our simulations with fiducial feedback strength and $\epsff=1\%$ (middle panel of Figure~\ref{fig:slopes}). This behavior is explained by the contribution of two comparable terms to $\tH2$ which can be demonstrated explicitly.

\begin{figure}
\centering
\includegraphics[width=\columnwidth]{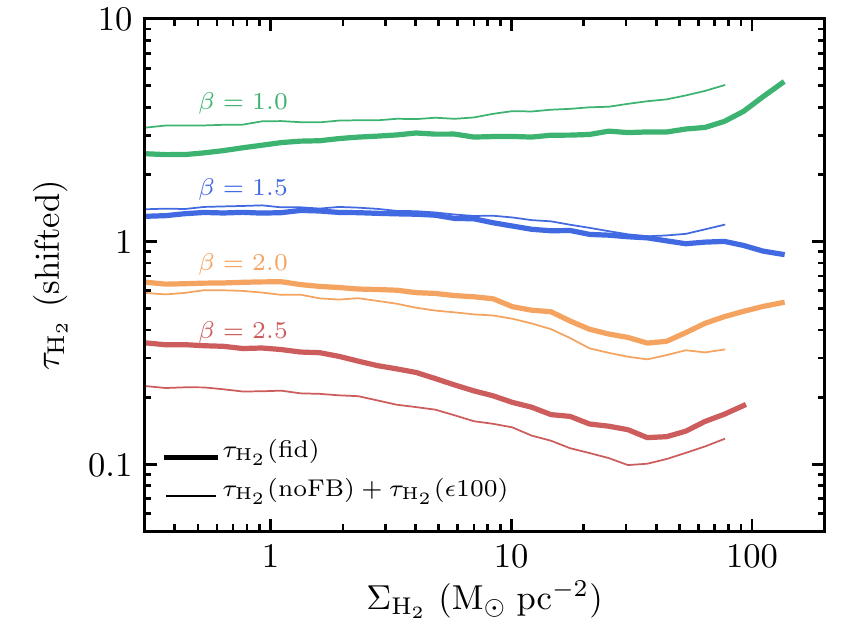}
\caption{\label{fig:ksr_sum} Comparison of our simulation results with the predictions of the analytical model. The thick lines show the $\tH2$ in the simulations with $\epsff=1\%$, fiducial feedback strength, and different slopes of the local star formation relation $\rhoSFR \propto \rho^\beta$, differentiated by color. These lines are equivalent to those shown in the middle panel of Figure~\ref{fig:slopes} but shifted up and down by an arbitrary factor to avoid clutter. Our model predicts that the $\tH2$ in these simulations must be close to the sum of the $\tH2$ in our simulations without feedback (left panel of Figure~\ref{fig:slopes}) and the $\tH2$ in our simulations with $\epsff=100\%$ (right panel of Figure~\ref{fig:slopes}). These sums are shown with thin lines for corresponding values of $\beta$ and they do indeed agree with the $\tH2$ in our fiducial simulations. }
\end{figure}

According to Equation~(\ref{eq:tauH2_sum}), the $\tH2$ in these simulations differ from the simulations with $\epsff=100\%$ and fiducial feedback only by the second term, which is 100 times smaller in the latter case. At the same time, $\tH2$ in the simulations with $\epsff=1\%$ and with or without feedback differ only by the first term. Therefore, Equation~(\ref{eq:tauH2_sum}) predicts that the $\tH2$ in our fiducial simulations equals the $\tH2$ in simulations with $\epsff=1\%$ and no feedback plus the $\tH2$ in simulations with $\epsff=100\%$ and fiducial feedback. 
Figure~\ref{fig:ksr_sum} shows that this is indeed the case. For $\beta=1.5$ and $2$, the measured $\tH2$ and the sum agree extremely well, while for $\beta=1$ and $2.5$, they are within a factor of $1.5$. A small difference in the latter case is due to the extra dependencies of $\tmolnsf$, $\tmd$, and $\taust$ on $\epsff$, $\beta$, and feedback strength. For example, feedback disperses the high-density tail of the gas PDF, making $\taust$ longer. This effect becomes stronger for larger $\beta$ because steeper local star formation relation $\rhoSFR \propto \rho^\beta$ results in a more efficient dispersal of dense gas. The difference between the measured $\tH2$ and the sum therefore increases with increasing $\beta$.

%--------------------------------------------------------------------
\subsection{Dependence on the Choice of Star Formation Threshold}
\label{sec:cycling:threshold}
%--------------------------------------------------------------------

\begin{figure}
\centering
\includegraphics[width=\columnwidth]{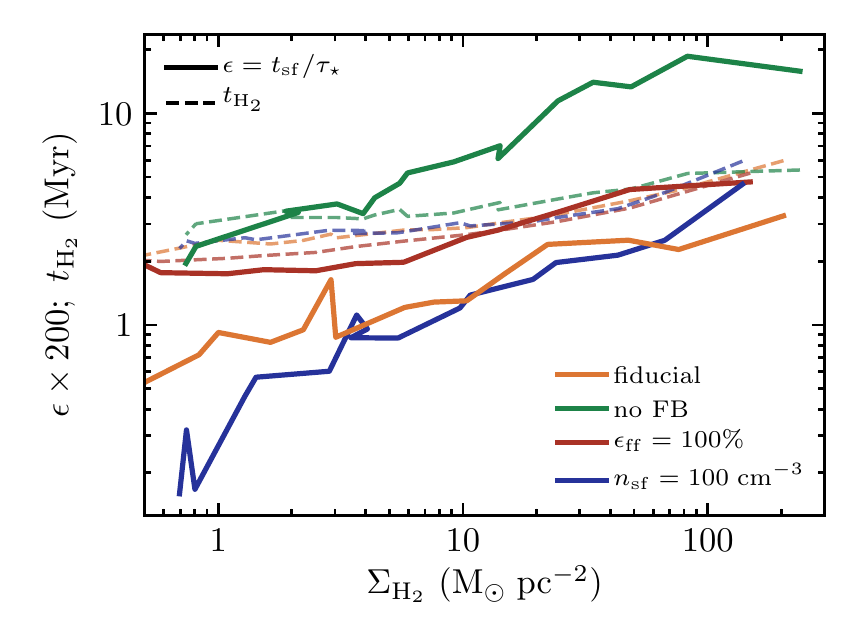}
\caption{\label{fig:tH2-eps} Effect of $\epsff$, feedback strength, and star formation threshold on the $\epsint(\SH2)$ and $\tmH2(\SH2)$ trends. The fiducial case shown by the orange lines corresponds to $\epsff=1\%$, fiducial feedback, and $\avirsf=10$ threshold. Other colors show simulations in which these parameters are changed as indicated in the legend. To illustrate the variation of trends with parameters, we compare simulations with the local slope $\beta=2$ instead of the fiducial $\beta=1.5$ because in these simulations trends vary more strongly due to a stronger dependence of $\taust$ on $\SH2$ (see Figures~\ref{fig:taust-eps} and \ref{fig:taust-eps-nofb}). The trend of $\epsint$ varies much stronger than that of $\tmH2$, and therefore, it is the trend of $\epsint$ that defines the dependence of $\tH2 \sim \tmH2/\epsint$ on $\SH2$ and the slope of the molecular KSR. }
\end{figure}

Simulation results in Section~\ref{sec:results} show that a near-linear KSR emerges only in simulations with efficient feedback {\it and} $\avir$-based star formation threshold. In simulations with the same parameters but using a density-based threshold, the molecular KSR is significantly steeper (see Figure~\ref{fig:ksr-n100}). This is not surprising, because the star formation threshold affects all terms in Equation~(\ref{eq:tauH2_sum}), and thus the behavior of the molecular KSR can be different for different threshold choices. Figure \ref{fig:tH2-eps} shows that $\tmH2(\SH2)$ is independent of the choice of threshold or feedback strength, but the $\epsint(\SH2)$ trend does steepen for the density-based star formation threshold and is thus responsible for the steepening of the molecular KSR in such simulation. This behavior can be understood using the gas distributions in simulations with different thresholds shown in Figure \ref{fig:phases}.
 
In all simulations, dense, supersonic gas reaches approximate equilibrium between turbulence production on the local dynamical time, $t_{\rm dyn} \propto 1/\sqrt{G \rho}$, and turbulence decay on the local eddy-turnover time, $t_{\rm dec} \sim \Delta/\st$, so that $t_{\rm dyn} \sim t_{\rm dec}$ and $\st \propto \rho^{0.5}$, and the gas PDF aligns along the lines of $\avir \propto \st^2/\rho \sim \const$, parallel to the $\avirsf=10$ threshold. This alignment persists at all $\SH2$.

In simulations with the $\avir$-based threshold, gas can become star-forming relatively quickly after becoming molecular because low-$\st$ gas can be star-forming even at rather small densities, while $\st$ can become small simply by turbulence decay when it cannot be offset with compression or feedback. In contrast, in simulations with the density-based threshold, transition to the star-forming state is hindered because molecular gas must be compressed to relatively high densities to exceed the threshold. 

\begin{figure}
\centering
\includegraphics[width=\columnwidth]{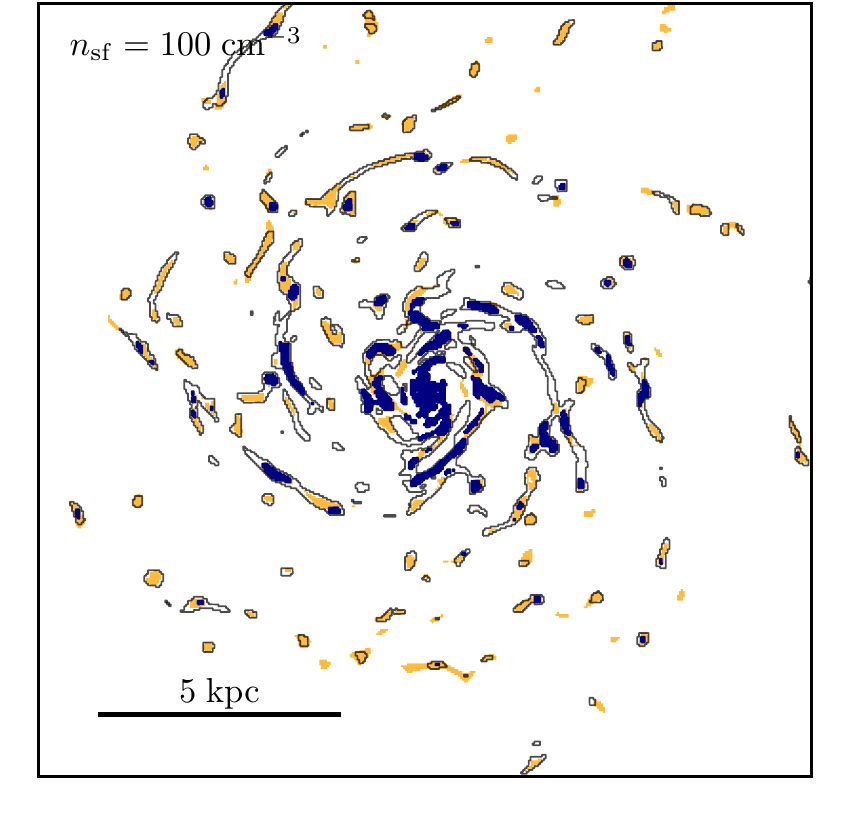}
\caption{\label{fig:n100_map2} Distribution of molecular (gray contours) and star-forming gas (filled blue contours) in the simulation with the $\nsf=100\cc$ star formation threshold. The filled orange contours show regions with $\avir<10$. Non-star-forming molecular regions on the disk outskirts in this simulation would be star-forming if the star formation threshold were in the virial parameter, $\avir < \avirsf = 10$. This is because it is much easier for gas to lose turbulence support and reach an $\avir$ threshold than to be compressed to a threshold density. }
\end{figure}

This difference can be illustrated using the distributions of $\SSFR$ and molecular gas in the middle panels of Figures~\ref{fig:ksr-avir10} and \ref{fig:ksr-n100}. In the simulation with the $\avirsf=10$ threshold (Figure~\ref{fig:ksr-avir10}), almost every molecular region (gray contours) contains a star-forming subregion with young stars, and thus molecular gas is always either star-forming or is spatially close to a star-forming region. In contrast, in the simulation with the density-based threshold (Figure~\ref{fig:ksr-n100}), a substantial number of molecular regions, especially in the disk outskirts, are not star-forming. As Figure~\ref{fig:n100_map2} shows, they would be star-forming
if we used the $\avir<10$ threshold instead of $n > 100\cc$.

The incidence of star-forming regions is thus a steeper function of $\SH2$ in the simulation with the density-based threshold and so is $\epsint(\SH2)$ (Figure \ref{fig:tH2-eps}), which explains the steeper molecular KSR. For a density-based threshold, the average $\epsint$ decreases at lower $\SH2$ because of the larger fraction of gas tracers that go through the molecular stage without reaching the star-forming state and thus have $\epsint=0$. 

Although we have considered specific choices of thresholds here, our conclusions and their implications are more general. The basic scaling of $\tH2$ with timescales in Equation~(\ref{eq:tauH2_tH2}) holds for any choice of star formation threshold, while as we can see, $\tmH2(\SH2)$ is insensitive to the details of feedback and star formation prescriptions. Thus, any prescription that shapes $\epsint(\SH2)$ to be similar to $\tmH2(\SH2)$ will lead to a near-linear molecular KSR and vice versa. Possible reasons why the $\epsint(\SH2)$ and $\tmH2(\SH2)$ trends are similar will be discussed in the next section. 

%--------------------------------------------------------------------
\subsection{The Origin of the Linear Molecular KSR}
\label{sec:linearKSR}
%--------------------------------------------------------------------

The results presented so far indicate that $\epsint$ and $\tmH2$ both depend on $\SH2$ and therefore, to produce a linear molecular KSR, these trends must cancel out in $\tH2 \sim \tmH2/\epsint$. According to Equation~(\ref{eq:tauH2_tH2}), $\taust/\fsf\sim\tmH2/\epsint$ and thus this cancellation is simply another manifestation of the cancellation of the $\taust(\SH2)$ and $\fsf(\SH2)$ trends discussed in Section~\ref{sec:pdf}.

We can readily  understand why the dependencies of $\epsint$ and $\tmH2$ on $\SH2$ should be of the same sign when feedback is efficient in dispersing star-forming regions. Indeed, $\epsint$ is expected to increase with increasing $\SH2$, because at higher $\SH2$, the gravity of the disk and pressure of the ISM are larger, making it harder for feedback to disperse star-forming regions and thus requiring a larger $\epsint$ for dispersal. At the same time, the time that the gas spends in the molecular state during one cycle, $\tmH2$, can also increase because a larger fraction of gas is molecular at larger $\SH2$.

The quantitative explanation of why these trends are similar is less obvious, but can be understood as follows. Our analysis shows that in simulations with efficient feedback, gas tracers experience {\em local} chaotic fluctuations of their density and velocity dispersion, and they move randomly in the \ns plane. In simulations with an $\avir$-based star formation threshold, most molecular gas is close to the threshold due to its alignment along the $\avir=\const$ direction (see Section~\ref{sec:cycling:threshold}), and therefore, it randomly transits into the star-forming state and back. If the probability of transition is close to uniform, the number of transitions during a molecular stage, $N_{\rm sf}$, will simply be proportional to the duration of this stage, $N_{\rm sf}\propto\tmH2$. Although the duration of each star-forming stage is regulated by feedback as we discussed above, the total time the parcel spends in the star-forming state, $\tsf$, will also be proportional to $N_{\rm sf}$ and the gas parcel thus converts the fraction $\epsilon\sim \tsf/\taust\propto\tmH2$ into stars during one molecular phase. This leads to a constant molecular depletion time, $\tH2\sim \tmH2/\epsilon\approx\rm const$, and a linear molecular KSR.

The above mechanism will also operate in a more general case of varying $\epsff$ as long as $\epsff$ is a strong function of $\avir$. Such a strong dependence of $\epsff$ on $\avir$ is a generic prediction of theoretical models of star formation in a turbulent medium \citep[see][for a review]{Padoan.etal.2014}. In Appendix \ref{app:turb} we show that the molecular KSR indeed remains linear when we vary $\epsff$ as an exponential function of $\avir$ instead of assuming a sharp threshold.

When feedback is inefficient, the evolution of molecular gas is not as chaotic, and the $\epsilon(\SH2)$ trend becomes sensitive to feedback strength and $\epsff$ value, as can be seen in Figure~\ref{fig:tH2-eps}. Given that the $\tmH2(\SH2)$ trend is nearly independent of feedback strength and $\epsff$, the depletion time $\tH2$ is no longer constant in such regimes.

Likewise, when a density threshold with a high value is used to define star-forming regions instead of $\avir$, most of the molecular gas is far from the threshold and the transition to a star-forming state in this case is not due to random motions of gas parcels in the \ns plane, but is mainly due to secular evolution and gas compression to high densities. In this case, $\tsf$ is no longer proportional to $\tmH2$ and $\tH2\ne\rm const$. For lower values of a density threshold, a larger fraction of gas is near the threshold and transitions to a star-forming state again become dominated by random fluctuations, which makes the slope shallower and closer to linear (see Figure~\ref{fig:app:thres} in the Appendix~\ref{app:sf-fb}). The slope in this case depends on the threshold value as the dynamical equilibrium between compression and turbulent pressure align gas along the $\avir=\rm const$ direction, not along $n=\rm const$. 

The presented explanation for why $\epsint$ and $\tmH2$ exhibit similar trends with $\SH2$, and thus why $\tH2\approx\rm const$ when feedback is strong, is admittedly qualitative. This question calls for further exploration both in high-resolution simulations of the ISM patches and observationally in studies of molecular and star-forming regions, as we discuss below in Section~\ref{sec:disc:obs}.

\vspace{2em}
%--------------------------------------------------------------------
\section{Discussion}
\label{sec:disc}
%--------------------------------------------------------------------

%--------------------------------------------------------------------
\subsection{Comparison to Previous Models}
\label{sec:disc:prev}
%--------------------------------------------------------------------

A commonly used explanation for the nearly linear molecular KSR is the so-called ``counting argument'' \citep[e.g.,][]{Wu.etal.2005}, in which one assumes that molecular regions have similar properties (e.g., density and thus depletion time $\taust$) and have a fixed fraction of gas that is undergoing star formation (i.e., $\fsf$). Then, $\tH2=\taust/\fsf$ (Equation~\ref{eq:tauH2_fsf}) becomes independent of $\SH2$ because both $\taust$ and $\fsf$ are the same and independent of $\SH2$. However, as we showed in Section~\ref{sec:sims-pdf}, our simulations indicate that both $\taust$ and $\fsf$ in molecular regions can vary with $\SH2$ and still produce a nearly linear molecular KSR. Furthermore,  properties of molecular clouds do change with galactocentric radius and thus with surface density in observed galaxies \citep[e.g.,][]{Heyer.Dame.2015,MivilleDeschenes.etal.2016}. The origin of the linear slope of the molecular KSR is therefore more nuanced.

\citet{Madore.2010} and later \citet{Elmegreen.2015,Elmegreen.2018} considered the origin of the KSR from the timescales of gas evolution in different states. Their approach is similar to the basis of our model, which allows a direct comparison. In particular, \citet{Madore.2010} expressed the depletion time of {\it total} gas as $\tau \equiv \Sg/\SSFR \sim (\tau_{\rm c} + \tau_{\rm s})/\epsilon$, where the duration of one gas cycle consists of the ``collapse time'' $\tau_{\rm c}$, on which average ISM gas evolves to the star-forming state, and ``stagnation time'' $\tau_{\rm s}$, on which star-forming gas is dispersed by feedback. The fraction of gas converted into stars in one cycle, $\epsilon$, is assumed to be fixed. Then, assuming also that $\tau_{\rm c} \propto \Sg^{-0.5}$ and $\tau_{\rm s}$ is constant, $\tau_{\rm s}$ will dominate at sufficiently high $\Sg$ and the KSR will become linear because the depletion time will become independent of $\Sg$: $\tau \sim \tau_{\rm s}/\epsilon$.

\citet{Elmegreen.2015,Elmegreen.2018} used principles similar to the \citet{Madore.2010} model but assuming that the {\it total} depletion time is proportional to the freefall time at the midplane density, $\tff$, with constant efficiency, $\epsff$, so that the depletion time of {\it molecular} gas\footnote{In \citet{Elmegreen.2015}, the molecular state is denoted by the subscripts ``CO,'' indicating that in observations, this state corresponds to the gas visible in CO. For consistency with our notation, we have changed subscripts ``CO'' to ``H$_2$.''} is $\tH2 = \fH2 \tff/\epsff$. Next, similarly to our Equation~(\ref{eq:fsf_tH2}), the molecular fraction is expressed as the ratio of timescales in the corresponding states: $\fH2 \sim t_{\rm ff,H_2}/(\tff+t_{\rm ff,H_2})$, where $t_{\rm ff,H_2}$ is the freefall time at the density of the molecular transition. The timescale $t_{\rm ff,H_2}$ is independent of $\SH2$ by construction because it is assumed to be set by the density of the atomic-to-molecular transition, which, in turn, is set by the local ISM properties rather than the large-scale surface density. For average ISM densities significantly lower than the molecular transition density, $\tff \gg t_{\rm ff,H_2}$ and therefore $\tH2 \sim t_{\rm ff,H_2}/\epsff$ becomes independent of $\SH2$ and the molecular KSR becomes linear. To compare with our model below, we note that in the \citet{Elmegreen.2015,Elmegreen.2018} model the duration of star formation cycle is assumed to be $\tff+t_{\rm ff,H_2}$ and thus the integral star formation efficiency per cycle is $\epsint = \epsff (\tff+t_{\rm ff,H_2})/\tff \sim \epsff$.

Although the models of \citet{Madore.2010} and \citet{Elmegreen.2015,Elmegreen.2018} are rather insightful, their prediction of the linear slope for the molecular KSR  follows from two strong assumptions that the characteristic time of molecular gas evolution ($\tau_{\rm s}$ in \citealt{Madore.2010} or $t_{\rm ff,H_2}$ in \citealt{Elmegreen.2015,Elmegreen.2018}) and the integral efficiency of star formation $\epsilon$ are all independent of the kiloparsec-scale gas surface density. Neither of these assumptions holds in our simulations. As was shown in Figure~\ref{fig:taust-eps}, $\epsint$ increases with $\SH2$ because at higher $\SH2$, feedback must overcome stronger forces to disperse a region and thus a larger fraction of molecular gas must be converted into stars. At the same time, the lifetime of molecular gas also increases. Nevertheless, in our simulations with the $\avirsf$ threshold and efficient feedback, the molecular KSR slope is still linear. As we showed, the slope is linear not because $\epsint$ and the lifetime of molecular gas are independent of $\SH2$, but because they scale with $\SH2$ similarly, and their trends nearly cancel.

Finally, \citet{Gnedin.etal.2014} argued that the linear molecular KSR on a $\gtrsim1\kpc$ scale is indicative of a linear relation on small scales (i.e., $\beta=1$ in our notation). As emphasized in that paper, this argument follows from the assumption that the gas PDF is self-similar {\it and} that star-forming gas can be defined using a threshold that is a simple function of gas density. In this case, the slope of the global relation is directly inherited from the small-scale relation in the regime when $\fsf \to 1$ (see ``special case 1'' in Appendix~\ref{app:pdf}). As we explicitly showed in Section~\ref{sec:sims-pdf}, the gas PDF in simulations with efficient feedback is not self-similar. This lack of self-similarity decouples the slope of the large-scale molecular KSR from the slope of the small-scale star formation recipe. Moreover, we showed that the definition of star-forming gas plays an important role in the resulting scaling, because the choice of the star-formation threshold explicitly enters the relevant equations (see Equations~\ref{eq:tauH2_fsf}--\ref{eq:sf-PDF}). Finally, our results clearly show that the variation of $\fsf$ with $\SH2$ is important, as it compensates the scaling of $\taust$ resulting from the particular scaling of the gas PDF with $\SH2$. 

%--------------------------------------------------------------------
\subsection{Implications for the Interpretations of the Observed Molecular KSR}
\label{sec:disc:obs}
%--------------------------------------------------------------------

The framework presented in Section~\ref{sec:ksr} is rather general and can help us elucidate the physical processes shaping the slope of the star formation relations observed in different galactic environments, on different spatial scales, and for different gas tracers. 

According to our model, a linear KSR for molecular gas is expected as long as the lifetime of gas in the molecular state, $\tmH2$, and the fraction of gas mass converted into stars over this lifetime, $\epsint$, exhibit similar trends with $\SH2$ so that these trends cancel in $\tH2 \sim \tmH2/\epsint$. 
The mechanism that makes the $\tmH2$ and $\epsint$ trends similar must be rather generic, because the KSR for molecular gas is observed to be linear in diverse environments: from the average ISM of normal star-forming galaxies across the Hubble sequence \citep{Utomo.etal.2017,Colombo.etal.2018} to low-metallicity dwarf galaxies \citep{Bolatto.etal.2011,Jameson.etal.2016} and low-density galactic disk outskirts \citep{Schruba.etal.2011}. 

An example of such a generic mechanism would be the efficient regulation of molecular stages by feedback. As we discussed above, feedback in our simulations makes evolution of molecular gas chaotic and the mass fraction of young stars formed by a given gas parcel during one molecular cycle becomes proportional to the duration of the cycle, $\epsilon\propto \tmH2$, which renders the molecular depletion time constant, $\tH2\sim \tmH2/\epsilon$. 

The molecular KSR steepens in high-density environments typical in starburst galaxies \citep[e.g.,][]{Genzel.etal.2010,Genzel.etal.2015,Leroy.etal.2013}. In the context of our model, the steepening corresponds to the regime in which feedback is less efficient in dispersing star-forming regions and dense molecular gas, the second term in Equation~(\ref{eq:tauH2_sum}) becomes comparable to the first, and $\epsilon$ is no longer proportional to $\tmH2$, which makes molecular KSR nonlinear. In addition, $\tmH2$ becomes shorter, due to either stronger gravity at higher gas surface densities or shorter turbulent crossing time at higher gas velocity dispersions.

Our model can also be used to interpret the KSR observed on scales much smaller than a kiloparsec and for gas states much denser than normal molecular gas. To this end, Equation~(\ref{eq:tauH2_tH2}) should be rewritten as
\begin{equation}
\label{eq:tauS}
\tau_S \sim \frac{\taust}{\tsf} t_S \sim \frac{t_S}{\epsilon},
\end{equation}
where $\tau_S$ is the depletion time of gas in a given state $S$, $t_S$ is the average time that a gas parcel spends in this state in a single cycle, and $\epsilon \sim \tsf/\taust$ is the fraction of gas converted into stars in one cycle through the state $S$. To apply this equation on different scales, its terms must be defined appropriately for the chosen scale.

For example, this equation can explain why the KSR becomes superlinear on the scales of individual star-forming regions, $\lesssim 50\pc$ \citep[e.g.,][]{Evans.etal.2009,Evans.etal.2014,Heiderman.etal.2010,Gutermuth.etal.2011}. Surveys of star-forming giant molecular clouds (GMCs) select only molecular gas from the star-forming stage, and therefore, the $t_S$ in Equation~(\ref{eq:tauS}) for such objects equals $\tsf$. According to Equation~(\ref{eq:tauS}), in this case depletion time is simply $\tau_{\rm GMC}\sim\taust$ and there is no longer feedback-imposed cancellation of the dependency in the $\taust/\tsf$ ratio on the large-scale ISM properties. In this picture, the slope of the small-scale KSR for star-forming regions is thus expected to reflect any dependence that $\taust$ has on the properties of these regions. Note also that mass fluxes of gas on the scales of GMCs are likely out of equilibrium, which will lead to a large variation of $\tH2$ \citep[see][for details]{Semenov.etal.2017} and thus a large scatter of the KSR, as is indeed observed on small scales \citep[e.g.,][]{Lee.etal.2016,Vutisalchavakul.etal.2016,Leroy.etal.2017}. Other sources of scatter include the intrinsic variation of gas properties in star-forming regions, incomplete sampling of different stages of gas evolution, and decoupled evolution of gas and star formation tracers \citep{Feldmann.etal.2011,Kruijssen.Longmore.2014}.

Equation~(\ref{eq:tauS}) can also be applied to interpret the linear relation between the amount of dense molecular gas traced by HCN and the total SFR observed on a wide range of scales \citep{Gao.Solomon.2004a,Gao.Solomon.2004b,Wu.etal.2005}. If we consider Equation~(\ref{eq:tauS}) on a $\lesssim 1\pc$ scale, typical for HCN gas, the star formation efficiency, $\epsilon$, will correspond to a fraction of a dense core mass that is eventually incorporated in a star \citep[e.g., $\epsilon \gtrsim 50\%$ in][]{Federrath.etal.2014}. The lifetime of such cores, $t_{\rm HCN}$, will be controlled by their local freefall time, on which the star is formed, and the rate at which the remaining dense gas is dispersed by feedback. With these definitions of $\epsilon$ and $t_{\rm HCN}$, the interpretation of the linear relation for HCN gas is similar to that for all molecular gas except that the separation between the scale on which HCN gas resides ($\lesssim 1\pc$) and the scale on which the relation is measured (up to the scale of entire galaxies) is much larger. Because of such large scale separation, both $\epsilon$ and $t_{\rm HCN}$ are expected to be independent of the large scale and thus $\tau_{\rm HCN} \sim t_{\rm HCN}/\epsilon$ will also be independent of the total HCN gas mass inside this scale, rendering the relation linear. Note, however, that the linear relation for dense molecular gas can also be in part an observational effect, due to the selection of gas from a narrow density range in the PDF tail \citep{Krumholz.Thompson.2007} or from the densest parts of isothermal spherical clouds \citep{Parmentier.2017}.

%--------------------------------------------------------------------
\subsection{Effect of Threshold on the Efficiency of Feedback in Galaxy Simulations}
\label{sec:disc:sims}
%--------------------------------------------------------------------

Our results show that the choice of star formation threshold in galaxy simulations has a significant impact on the resulting slope of the KSR. Contrary to previous arguments in the literature, we show that in our simulations the slope of the molecular KSR does not merely reflect the density dependence of the star formation recipe adopted at the resolution scale, but depends on the density PDF of the ISM gas, which is strongly affected by feedback. Physically, the threshold determines both the locations where current star formation proceeds {\em and} the timescale and efficacy with which feedback can render star-forming gas non-star-forming. The strong effect that the choice of threshold has on the results implies that this choice must be made with great care, as was indeed demonstrated in related contexts by \citet{Governato.etal.2010} and \citet{Hopkins.etal.2013}.  

As was shown in Sections~\ref{sec:cycling:fb}--\ref{sec:linearKSR}, the role of feedback in making the KSR linear is twofold. First, feedback must efficiently disperse star-forming gas, which results in self-regulation to constant star formation efficiency per cycle, $\epsilon$, independent of the rate at which gas is converted into stars in the star-forming state. Second, feedback must be efficient in converting molecular gas into the atomic state and establishing a correlation between $\tmH2$ and $\epsint$ which leads to a near cancellation of their trends with $\SH2$ in $\tH2 \sim \tmH2/\epsint$ and thus to a near-linear KSR.

Our results indicate that the efficacy of stellar feedback in both of these aspects is much higher when the star formation threshold is based on $\avir$, compared to the threshold based on constant gas density. Indeed, it is generally faster to render gas non-star-forming by driving subgrid turbulence and thereby increasing $\avir$ than to actually disperse a star-forming region and decrease its density. Similarly, the lifetime of gas in the molecular phase is also controlled by feedback to a much larger degree, because gas becomes star-forming shortly after it becomes molecular and therefore has less time to become denser and is more difficult to be dispersed.

The above arguments and our simulation results thus favor a star formation threshold in virial parameter \citep[see also][]{Hopkins.etal.2013} rather than the more popular threshold in gas density. Not only is such a threshold well motivated by models of star formation in turbulent ISM, it also naturally leads to a linear KSR for molecular gas when feedback is efficient. The use of this threshold in practice is somewhat complicated by the necessity to estimate turbulent velocity dispersion at the resolution scale and generally requires modeling of turbulent velocities on subgrid scales. On the other hand, our results indicate that simulations that use density-based or the $f_{\rm H_2}$ star formation threshold should be interpreted with caution, especially on the sub-galactic scale, where the gas distribution is strongly affected by the choice of threshold.

The optimal choice of the threshold and other parameters of star formation and feedback models will of course depend on the scale on which these models are applied. In particular, the mechanism of the linear KSR origin on kiloparsec scales requires high resolution so that the transitions of gas between atomic, molecular, and star-forming states on sub-kiloparsec scales are sufficiently resolved. The resolution of state-of-the-art large-volume cosmological simulations, $\lesssim1\kpc$, is not yet sufficient to capture these transitions, and thus, the slope of the KSR on kiloparsec scales reflects that adopted in the star formation prescription. 

%--------------------------------------------------------------------
\section{Summary and conclusions}
\label{sec:summary}
%--------------------------------------------------------------------

Using a suite of isolated \Lstar~galaxy simulations, we explored the origin of the slope of the relation between surface densities of molecular gas and SFR averaged on kiloparsec scales. We showed that when feedback is efficient and the star formation threshold is based on the virial parameter, this relation has a near-linear slope, regardless of the slope adopted in the resolution-scale relation between star formation rate and gas density. Thus, in this regime, the slope of the KSR on kiloparsec scales does not reflect the slope on small scales, contrary to a number of previous arguments in the literature. 

We showed that the linear slope of the molecular KSR and its insensitivity to the local slope result from the particular scaling of the gas PDF with the gas surface density. When feedback is efficient in shaping the PDF, the PDF scaling leads to a cancellation of trends exhibited by the average depletion times in star-forming gas and by star-forming mass fractions. When feedback is not efficient, the gas PDF is shaped by dynamical processes, such as ISM turbulence and passage of spiral arms, and the KSR becomes dependent on the local slope and thus is not necessarily linear.

We explained these results using an analytical model based on the conservation of interstellar gas mass as the gas cycles between atomic, molecular, and star-forming states (see Section~\ref{sec:ksr})---the model we previously used to explain the physical origin of the gas depletion time and its dependence on star formation efficiency and feedback strength \citep{Semenov.etal.2017,Semenov.etal.2018}. Our main findings can be summarized as follows:

\begin{enumerate}
\item In Section~\ref{sec:cycling:fb}, we show that when feedback is efficient in dispersing star-forming gas, the typical duration of star-forming stages, $\tsf$, is proportional to the local depletion time of star-forming gas, $\taust$, so that the molecular depletion time, $\tH2 \propto \tmH2\,\taust/\tsf$, becomes independent of $\taust$. This explains why the molecular KSR is insensitive to the local slope adopted in the subgrid prescription for star formation, because only $\taust$ depends explicitly on the local slope (see Equation~\ref{eq:taust-PDF}), while the time the gas spends in molecular form during each cycle, $\tmH2$, does not.  

\item When feedback is inefficient, $\tsf$ is controlled by dynamical processes that disperse star-forming gas: turbulent shear, differential rotation of galactic disk, expansion behind spiral arms, etc. Star formation, and thus $\taust$, reflects the gas PDF shaped by these dynamical processes, but does not affect this PDF via feedback. Thus, the trend of $\tsf$ is no longer proportional to $\taust$, and the molecular KSR becomes dependent on the local slope of the star formation prescription $\beta$. 

\item Simulations in the regime intermediate between the regimes of efficient and inefficient feedback exhibit intermediate behavior. Thus, in our fiducial simulation with $\epsff=1\%$, the effects of feedback and dynamical processes are both important, and therefore, the molecular KSR is close to linear, but its slope weakly depends on the local slope (see Figure~\ref{fig:slopes} and Section~\ref{sec:cycling:fb}).

\item We show that a near-linear molecular KSR emerges only in simulations that use a star formation threshold in the virial parameter. The molecular KSR generally has a nonlinear slope in simulations with the same efficiency and feedback strength, but that use a density-based threshold, with the slope steepening with the increasing threshold value. We argue that this is because in the latter simulations, the time between the moment gas becomes molecular and the moment it becomes star-forming is much longer. As a result, at lower $\SH2$, a smaller fraction of molecular gas is able to reach the star-forming state, and the molecular KSR therefore becomes steep.
\end{enumerate}

The theoretical framework we use to interpret our simulation results is rather general and can be used to interpret observations as well. For example, it can shed light on the origin of star formation relations observed for different gas tracers, on different spatial scales, and in different galactic environments, as we discuss in  Section~\ref{sec:disc:obs}.

The framework is also useful for interpreting and designing galaxy formation simulations. In particular, our results indicate that attention should be paid not only to the modeling of feedback but also to the modeling of star formation and, in particular, the choice of criteria used to identify star-forming gas. Our simulation results favor a criterion based on the local virial parameter instead of the commonly used density-based criterion (Section~\ref{sec:disc:sims}).

\acknowledgements
We thank the anonymous referee for constructive feedback and valuable suggestions.
A.K. and N.G. would like to thank participants and organizers of the Simons symposium series on galactic superwinds for stimulating discussions that played a role in motivating this study. 
This work was supported by a NASA ATP grant NNH12ZDA001N, NSF grants AST-1412107 and AST-1714658, and by the Kavli Institute for Cosmological Physics at the University of Chicago through grant PHY-1125897 and an endowment from the Kavli Foundation and its founder, Fred Kavli. The simulation and analyses presented in this paper have been carried out using the Midway cluster at the University of Chicago Research Computing Center, which we acknowledge for support. Analyses presented in this paper were greatly aided by the following free software packages: {\tt yt} \citep{yt}, {\tt NumPy} \citep{numpy_ndarray}, {\tt SciPy} \citep{scipy}, {\tt Matplotlib} \citep{matplotlib}, and \href{https://github.com/}{GitHub}. We have also used the Astrophysics Data Service (\href{http://adsabs.harvard.edu/abstract_service.html}{ADS}) and \href{https://arxiv.org}{arXiv} preprint repository extensively during this project and writing of the paper.

\appendix

%--------------------------------------------------------------------
\section{Dependence of the molecular KSR slope on local star formation and feedback parameters}
\label{app:sf-fb}
%--------------------------------------------------------------------

\begin{figure}
\centering
\includegraphics[width=\columnwidth]{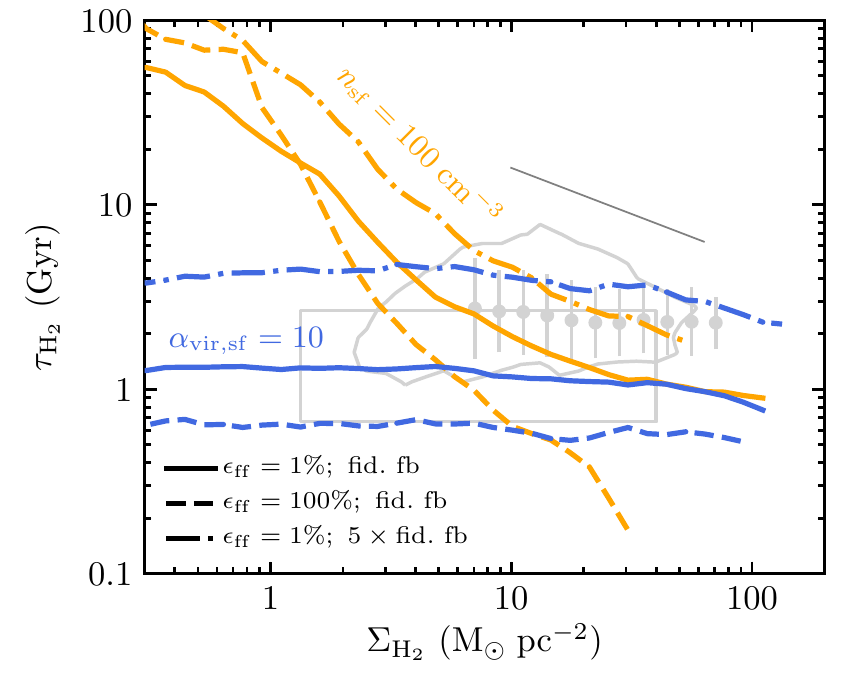}
\caption{\label{fig:app:epsff-fb} Median depletion time of the molecular gas as a function of $\SH2$ in simulations with $\avirsf=10$ (blue lines) and $\nsf = 100\cc$ star formation thresholds (orange lines), and different values of local star formation efficiency $\epsff$ and feedback strength. Solid lines show $\tH2$ for fiducial feedback strength and $\epsff=1\%$ (i.e., the same as in Figures~\ref{fig:ksr-avir10} and \ref{fig:ksr-n100}), dashed lines show $\tH2$ for higher $\epsff=100\%$, and dashed-dotted line show $\tH2$ in simulations where the fiducial momentum input from supernovae is multiplied by a factor of 5 (see Section~\ref{sec:sims}). Gray contours and points with error bars indicate the observed $\tH2$ (see the legend in Figures~\ref{fig:ksr-avir10}), and the thin gray line shows the slope adopted in the star formation prescription: $\rhoSFR = \epsff\, \rho/\tff \propto \rho^{1.5}$ and thus $\rho/\rhoSFR \propto \rho^{-0.5}$. }
\end{figure}

Figure~\ref{fig:app:epsff-fb} shows the dependencies of the molecular KSR slope on the efficiency of star formation per freefall time, $\epsff$, and feedback strength in our simulations with $\avirsf=10$ (blue lines) and $\nsf=100\cc$ star formation thresholds (orange lines). In simulations with the $\avirsf$ threshold, the slope remains linear and only the normalization changes at different $\epsff$ and feedback strength, as we showed in \citet[][see Figure~9 and related discussion]{Semenov.etal.2018}. In simulations with the $\nsf$ threshold, in contrast, the slope becomes steeper for larger values of $\epsff$.

\begin{figure}
\centering
\includegraphics[width=\columnwidth]{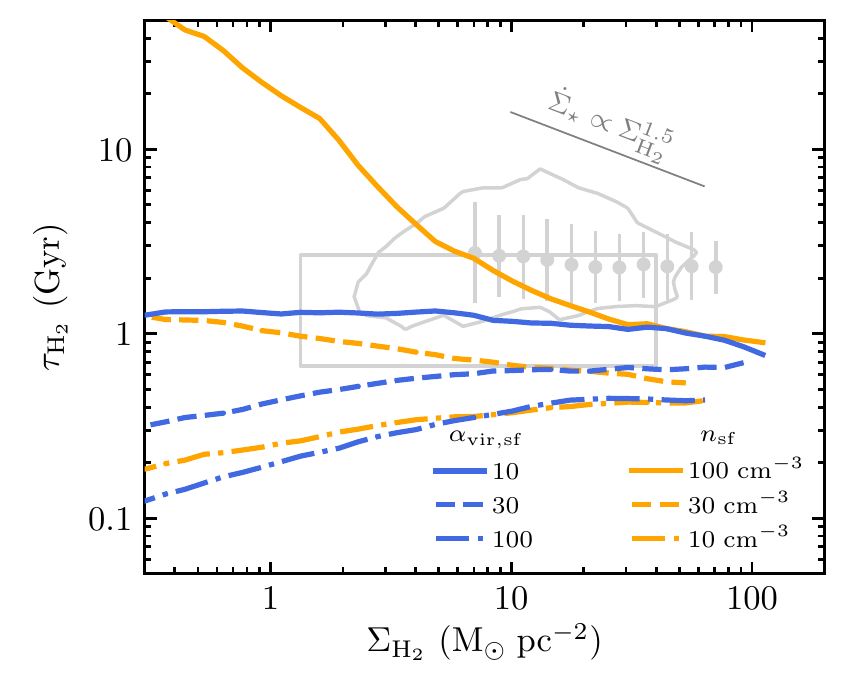}
\caption{\label{fig:app:thres} Same as Figure~\ref{fig:app:epsff-fb} but for different values of the star formation threshold: $\avirsf = 10$, 30, and 100 (blue lines), and $\nsf = 100$, 30, and $10\cc$ (orange lines). }
\end{figure}

Figure~\ref{fig:app:thres} shows the change of the molecular KSR slope in simulations with different values of the $\avir$ and density threshold. For an $\nsf$ threshold, the slope becomes shallower for lower $\nsf$: for $\nsf \gtrsim 30\cc$, the slope is steeper than linear, but it becomes shallower than linear for smaller $\nsf$ because non-molecular gas is identified as star-forming. For the $\avirsf$ thresholds, the sensitivity of the slope to the $\avirsf$ value is much weaker, although the slope still becomes somewhat shallower for very large values of $\avir$.

%--------------------------------------------------------------------
\section{Molecular KSR slope in simulations with explicitly modeled $\epsff$}
\label{app:turb}
%--------------------------------------------------------------------

\begin{figure}
\centering
\includegraphics[width=\columnwidth]{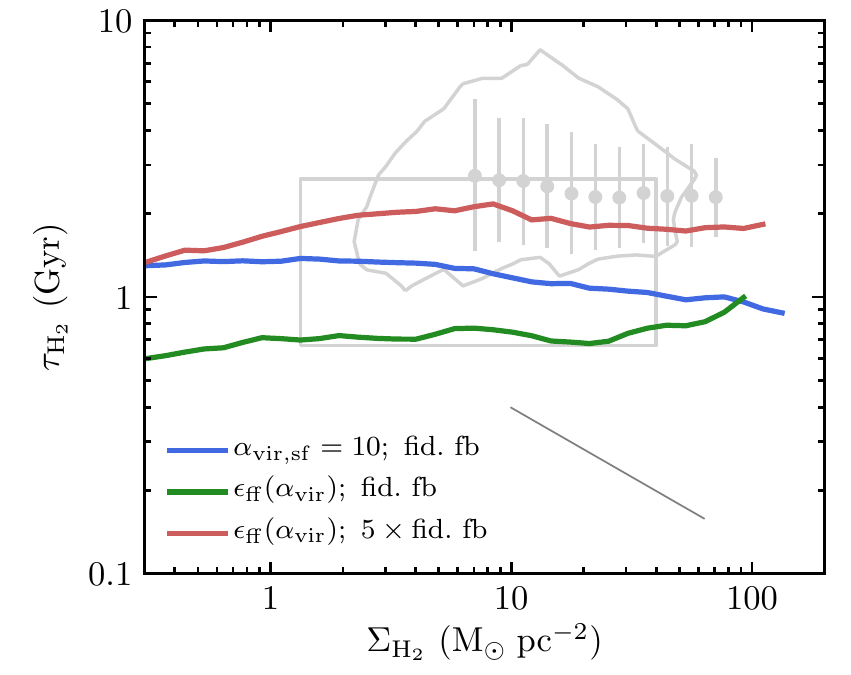}
\caption{\label{fig:app:turb} Median depletion time of molecular gas as a function of $\SH2$ in simulations with $\epsff$ continuously varied according to Equation~(\ref{eq:app:PHN12}). Green and red lines show the results for our fiducial and 5 times stronger feedback, respectively. For reference, the blue line shows the results of our fiducial simulation with $\epsff=1\%$ in gas defined by a sharp star formation threshold, $\avir<10$. The thin gray line shows the slope of $\tH2 \propto \SH2^{-0.5}$.}
\end{figure}

As was shown in Figure~\ref{fig:ksr-avir10}, our fiducial simulation with the $\avirsf=10$ star formation threshold and constant $\epsff=1\%$ reproduces the observed near-linear slope of the molecular KSR. Figure~\ref{fig:app:turb} shows that the molecular KSR remains linear when we  vary $\epsff$ using the fit to the simulation results of \citet{Padoan.etal.2012}:
\begin{equation}
\label{eq:app:PHN12}
\epsff = \exp( -\sqrt{ \avir/0.53 } ).
\end{equation}
The figure also shows that the normalization of the KSR is affected by feedback strength in the same way as in the simulations with a sharp $\avir$ threshold (see Appendix~\ref{app:sf-fb}).

The KSR in simulations with varied $\epsff$ is similar to that in our fiducial simulation because our threshold choice, $\avirsf=10$, approximates the exponential increase of $\epsff$ for $\avir\lesssim 10$ in Equation~(\ref{eq:app:PHN12}). A factor of $\sim 2$ difference in normalization can be explained by the somewhat higher $\epsff$ values predicted by Equation~(\ref{eq:app:PHN12}), compared to our fiducial $\epsff=1\%$.
Indeed, the $\avir$ of the star-forming gas in our fiducial simulation ranges between 10 and $\sim 2$ (see Figure~\ref{fig:phases}), and Equation~(\ref{eq:app:PHN12}) predicts $\epsff \sim 1.3\%\text{--}14\%$ for such $\avir$.

Interestingly, \citet{Lupi.etal.2018} used a star formation prescription with varying $\epsff$ and also found a shallow molecular KSR. Their KSR is somewhat steeper than linear, which can be due to a strong adopted dependence of $\epsff$ on density (see their Appendix~A) and the resulting large effective local slope $\beta$. This result is consistent with our fiducial simulations with high $\beta$ that also have a somewhat steeper than linear molecular KSR (see Figure~\ref{fig:slopes}). \citet{Lupi.etal.2018} also found that the molecular KSR slope depends on the assumptions about the unresolved clumping factor of the gas.

%--------------------------------------------------------------------
\section{Scale dependence of the molecular KSR slope}
\label{app:scale}
%--------------------------------------------------------------------

\begin{figure}
\centering
\includegraphics[width=\columnwidth]{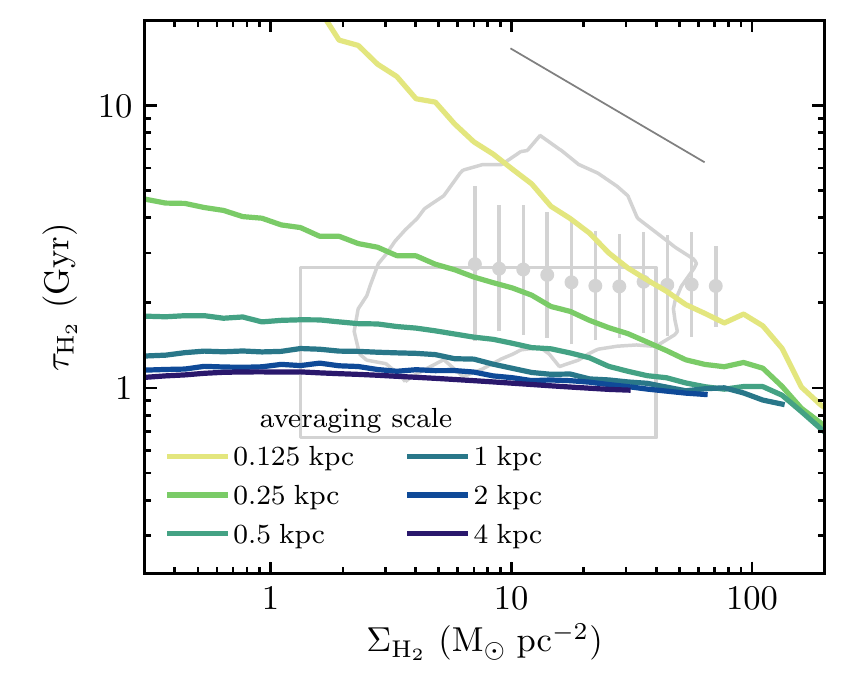}
\caption{\label{fig:app:scale} The dependence of median $\tH2=\SH2/\SSFR$ on the width of the 2D Gaussian filter used to average the $\SSFR$ and $\SH2$ maps in our fiducial simulation with the $\avirsf=10$ star formation threshold and $\epsff=1\%$. When the averaging scale is close to the resolution scale ($40\pc$), the molecular KSR slope approaches that adopted in the star formation prescription (shown by the gray line) and can become somewhat steeper due to the large scatter of $\tH2$ on these small scales. At $\sim 500\pc$, the slope becomes near-linear (i.e. $\tH2 \approx \const$), and it stays linear at any larger scale.}
\end{figure}

In this work, we focused on the molecular KSR averaged on $1\kpc$ scales. The scale dependence of the KSR is an interesting related topic \citep[e.g.,][]{Feldmann.etal.2011,Khoperskov.Vasiliev.2017,Orr.etal.2018}. Figure~\ref{fig:app:scale} shows how the slope of the molecular KSR in our fiducial simulation depends on the averaging scale. As expected, on scales close to the resolution scale of our simulation ($40\pc$), the KSR slope approaches the slope of the adopted star formation prescription, $\beta=1.5$. However, as the averaging scale increases, the molecular KSR flattens and becomes near-linear at $\gtrsim 500\pc$ scales. At any larger scale, the molecular KSR remains linear because when the slope is linear for a certain smoothing scale, it always remains linear on larger scales, as discussed for {\it special case 2} in Appendix~\ref{app:pdf}. 

The transition scale of $\sim 500\pc$ reflects the spatial coherence of star-forming and non-star-forming molecular gas in our simulations. Patches of $<500\pc$ size preferentially include only one of the states, while larger patches are sufficient to average between both states. This effect also leads to a scatter that increases on smaller scales \citep{Kruijssen.Longmore.2014}. A similar coherence scale was also obtained for M33 by \citet{Schruba.etal.2011}.  

%--------------------------------------------------------------------
\section{Derivation of the equations connecting the molecular KSR with gas PDF and star formation on small scales}
\label{app:pdf}
%--------------------------------------------------------------------

The depletion time of molecular gas in a single kiloparsec-scale patch can be expressed as a function of the local SFR density, $\rhoSFR$, and molecular gas mass in the patch, $M_{\rm H_2}$, as follows:
\begin{equation}
\label{eq:app:tauH2-dV}
\frac{1}{\tH2} = \frac{\SSFR}{\SH2} = \frac{\dot{M}_{\star}}{M_{\rm H_2}}= \frac{1}{M_{\rm H_2}} \int \rhoSFR dV,
\end{equation}
where integration is carried out over the patch volume.  If we assume that $\rhoSFR$ is a power-law function of gas density, we can 
write $\rhoSFR$ as 
\begin{equation}
\rhoSFR = A \rho^\beta\,\Theta_{\rm sf}(\rho,\q),
\end{equation}
where $\Theta_{\rm sf}(\rho,\q)$ is a function varying from $0$ to $1$ that defines star-forming gas as a function of gas density and a vector of other relevant properties $\q$. Furthermore, we can express the volume element $dV$ as
\begin{equation}
\label{eq:app:dV}
dV = \frac{\partial^2 V}{\partial \rho \partial \q} d\rho d\q = \frac{M_{\rm H_2}}{\fH2 \rho} P_{\rm H_2}(\rho,\q) d\rho d\q,
\end{equation}
where
\begin{equation}
P_{\rm H_2}(\rho,\q) = \frac{1}{M_{\rm H_2}} \frac{\partial^2 M_{\rm H_2}}{\partial \rho \partial \q} = \frac{\fH2 \rho}{M_{\rm H_2}} \frac{\partial^2 V}{\partial \rho \partial \q}
\end{equation}
is the mass-weighted distribution of molecular gas, and $\fH2$ is the local mass fraction of molecular gas, which can be a function of total gas density $\rho$, metallicity, radiation field, and other properties. In general, the variation of $\fH2$ in star-forming regions can be accounted for. However, most of the star-forming gas selected by our criteria has $\fH2 \approx 1$, and therefore, we will adopt $\fH2 = 1$ {\em in star-forming gas}. Under this assumption, Equations~(\ref{eq:app:tauH2-dV})--(\ref{eq:app:dV}) can be combined to
\begin{equation}
\label{eq:app:tauH2-patch}
\frac{1}{\tH2} = A \iint_0^{\infty} \rho^{\beta-1} \Theta_{\rm sf}(\rho,\q) P_{\rm H_2}(\rho,\q) d\rho d\q.
\end{equation}

Equation~(\ref{eq:app:tauH2-patch}) describes the relation between molecular gas depletion time and the PDF of molecular gas in a single ISM patch. The shape of $P_{\rm H_2}(\rho,\q)$ in this equation can vary from patch to patch, leading to variation of $\tH2$. To obtain the molecular KSR, this equation must be averaged between patches with the same molecular surface density $\SH2$:
\begin{equation}
\label{eq:app:tauH2-ave}
\left\langle \frac{1}{\tH2} \right\rangle = A \iint_0^{\infty} \rho^{\beta-1} \Theta_{\rm sf}(\rho,\q) \langle P_{\rm H_2} \rangle (\rho,\q | \SH2) d\rho d\q,
\end{equation}
where the average shape of $\langle P_{\rm H_2} \rangle$ will depend on $\SH2$ and this dependence will define the dependence of the average $\tH2$ on $\SH2$, i.e. the slope of the molecular KSR.

Finally, omitting explicit averaging to simplify notation, Equation~(\ref{eq:app:tauH2-ave}) can be rewritten as
\begin{equation}
\label{eq:app:tauH2-taust}
\frac{1}{\tH2} = \frac{\fsf}{\taust},
\end{equation}
where, by definition,
\begin{equation}
\label{eq:app:fsf}
\fsf = \iint_0^{\infty} \Theta_{\rm sf}(\rho,\q) P_{\rm H_2} (\rho,\q | \SH2) d\rho d\q
\end{equation}
is the star-forming mass fraction of molecular gas and
\begin{equation}
\label{eq:app:taust}
\frac{1}{\taust} = A \int_0^{\infty} \rho^{\beta-1} P_{\rm sf} (\rho | \SH2) d\rho
\end{equation}
is the inverse local depletion time $1/\tdep = \rhoSFR/\rho = A\rho^{\beta-1}$ averaged over the density PDF of star-forming gas:
\begin{equation}
\label{eq:app:psf}
P_{\rm sf}(\rho|\SH2) = \frac{\int \Theta_{\rm sf}(\rho,\q) P_{\rm H_2}(\rho,\q|\SH2) d\q}{\iint_0^{\infty} \Theta_{\rm sf}(\rho,\q) P_{\rm H_2}(\rho,\q|\SH2) d\rho d\q}.
\end{equation}

Equations~(\ref{eq:app:tauH2-taust}--\ref{eq:app:psf}) show that the connection between the slope of the $\SSFR$--$\SH2$ relation on kiloparsec scales and its local value $\beta$ is nontrivial and in general depends on the scaling of gas PDF with $\SH2$. However, in some special cases studied previously in the literature, these equations predict a direct relation between global and local slopes. In these special cases, $\fsf$ is assumed to be independent of $\SH2$ as would be the case if, e.g., all molecular gas were star-forming, $\fsf = 1$. The KSR slope is then determined only by the behavior of the integral in the definition of $\taust$ (Equation~\ref{eq:app:taust}). 

{\em Special case 1.} Star-forming gas PDF scales self-similarly with $\SH2$: $P_{\rm sf}(\rho | \SH2) = F(\rho/\SH2)/\SH2$. In this case, $\taust$ inherits the dependence on $\SH2$ from the local star formation relation, $1/\tdep \propto \rho^{\beta-1}$:
\begin{equation}
\frac{1}{\taust} = A \SH2^{\beta-1} \int_0^\infty x^{\beta-1} F(x) dx \propto \SH2^{\beta-1},
\end{equation}
and therefore the KSR inherits the local slope $\beta$: $\SSFR \propto \SH2^\beta$ \citep[cf.][]{Gnedin.etal.2014}.

{\em Special case 2.} Local relation is linear, $\beta=1$. In this case,
\begin{equation}
\frac{1}{\taust} = A \int_0^\infty P_{\rm sf}(\rho | \SH2) d\rho = A,
\end{equation}
and the KSR is also linear: $\SSFR = A \fsf \SH2$ \citep[cf.][]{Gnedin.etal.2014}. Physically, when $\beta=1$, the local depletion time $\tdep = \rho/\rhoSFR = \rho^{1-\beta}/A = A^{-1}$ is constant in all star-forming gas and thus its average $\taust = \tdep = A^{-1}$ is independent of $\SH2$. This means, for example, that the linear molecular KSR observed on kiloparsec scales will remain linear when averaged on any larger scale (see Figure~\ref{fig:app:scale}).

{\em Special case 3.} The shape of the star-forming gas PDF is independent of $\SH2$: $P_{\rm sf}(\rho | \SH2) = F(\rho)$ and  
\begin{equation}
\frac{1}{\taust} = A \int_0^\infty \rho^{\beta-1} F(\rho) d\rho
\end{equation}
becomes independent of $\SH2$, and therefore the KSR becomes linear regardless of the local slope $\beta$: $\SSFR \propto \SH2$. This is a more general case of the ``counting argument,'' in which all star-forming regions are assumed to have the same density $\rho_0$ so that $P_{\rm sf}(\rho | \SH2) = \delta(\rho-\rho_0)$.

\bibliographystyle{aasjournal}
\bibliography{}

\end{document}